\documentclass{elsarticle}
\makeatletter
\def\ps@pprintTitle{ \let\@oddhead\@empty
 \let\@evenhead\@empty
 \def\@oddfoot{\centerline{\thepage}} \let\@evenfoot\@oddfoot}
\makeatother

\usepackage{lineno}

\usepackage[table]{xcolor}
\usepackage{graphicx}\usepackage{rotating}\usepackage{pict2e}

\usepackage{dcolumn}\usepackage{bm}                          \usepackage[hidelinks]{hyperref} \usepackage[all]{hypcap}                         

\usepackage{relsize}

\usepackage{booktabs}

\usepackage{mathtools}

\newcommand{\enquote}[1]{``#1''} 
\usepackage{xspace}

\newcommand{\ev}{{\rm e}\kern-1.pt{\rm V}}
\newcommand{\gev}{{\rm Ge}\kern-1.pt{\rm V}}
\newcommand{\mev}{{\rm Me}\kern-1.pt{\rm V}}
\newcommand{\kev}{{\rm ke}\kern-1.pt{\rm V}}
\newcommand{\tev}{{\rm Te}\kern-1.pt{\rm V}}
\newcommand{\gevsq}{\mbox{$\mathrm{{\rm Ge}\kern-1.pt{\rm V}}^2$}}

\newcommand{\degree}{\mbox{\kern-1.pt$^\circ$}\xspace}

\def\lsim{\mathrel{\rlap{\lower4pt\hbox{\hskip1pt$\sim$}}
    \raise2pt\hbox{$<$}}} 
\def\gsim{\mathrel{\rlap{\lower4pt\hbox{\hskip1pt$\sim$}}
    \raise2pt\hbox{$>$}}}

\begin{document}

\title
{
Error Determination in Sextupole Magnet Calibration and Alignment Measurements and Application to Horizontal Beam Size Calculations at the Cornell Electron-positron Storage Ring
}

\author{J.A.~Crittenden}
\ead{crittenden@cornell.edu}
\author{G.H.~Hoffstaetter}
\author{D.C.~Sagan}

\address[Cornell]{CLASSE\fnref{fn1}, Cornell University, Ithaca, NY 14853, United States}

\begin{abstract}
We report on measurements and modeling studies performed from 2021 to 2024 on the 76 sextupole magnets in the Cornell Electron-positron Storage Ring CESR. Beam-based, magnet-specific calibrations ($K_2$ value versus excitation current) were measured, replacing the common value obtained from transverse field measurements and longitudinal modeling from the late 1990's. The method consists of defining a custom closed bump for each sextupole and measuring the slope of the betatron tune change as a function of horizontal beam position. It was found that the new calibrations differ by an average of 3.1\% with an RMS spread of 12\%. The uncertainties in the calibration correction factors average 1.7\% with an RMS spread of 1.0\%.

Sextupole alignment values relative to the reference orbit were measured by combining the measured beam position with the quadrupole and skew quadrupole terms caused by a sextupole strength change $\Delta K_2$. High accuracy was achieved by fitting to difference phase and coupling functions as $K_2$ was varied. The horizontal (vertical) average offset values were found to be -0.01 (0.03)~mm with RMS spread of 1.1 (0.9)~mm  with some exceptionally large values of a few millimeters.
Typical uncertainties are 0.01-0.02~mm.

The above measurements were motivated by the precision required in measuring horizontal beam size at each sextupole. A precision of 10\%  for a \mbox{1-mm} beam size requires uncertainties of better than 0.1~$\mu$radian in the horizontal angle change produced in the sextupole for a typical strength change of \mbox{$\Delta K_2 \, L$ = 1 m$^{-2}$}, where $L$ is the length of the sextupole, as well as 10\% in the difference of the squared horizontal and vertical beam positions relative to the center of the sextupole. These precision values were achieved by the analysis of difference functions. However, a small source of horizontal angle change of unknown origin, independent of the sextupole strength, requires a sextupole strength range larger than now available to measure accurately the typical horizontal beam size at CESR.

\end{abstract}

\maketitle

\vspace*{-2cm}
{\hfill Draft of \today}

\tableofcontents

\section{Introduction}
\label{sec:introduction}
The sextupole field components \mbox{$\frac{qL}{p_0}B_{\rm X} =  K_2L xy$} and \mbox{$\frac{qL}{p_0}B_{\rm Y} = \frac{1}{2} K_2L (x^2 - y^2 )$} can be used to derive expressions for the quadrupole kick $\Delta  b_1$, the skew quadrupole kick $\Delta  a_1$ and the orbit angle changes $\Delta  p_{\rm X}$ and $\Delta  p_{\rm Y}$ caused by a change in sextupole strength $\Delta K_2L$ as follows. Assuming initial $K_2 = 0$ and including the parabolic and  cubic terms,
   \begin{eqnarray}
     \Delta  b_1 &=& \Delta K_2L \left( X_0 + \Delta {\rm x} \right) \label{eq:db1}\\
     \Delta  a_1 &=& \Delta K_2L \left( Y_0 + \Delta {\rm y} \right) \label{eq:da1}\\
     \Delta  p_{\rm Y} &=& \Delta K_2L \left( X_0 + \Delta {\rm x} \right) \left( Y_0 + \Delta {\rm y} \right) \label{eq:dpy}
   \end{eqnarray} 
   \begin{eqnarray}
    \resizebox{0.9\hsize}{!}{$    	 \Delta  p_{\rm X}=\frac{1}{2} \Delta K_2L \left[  \left( Y_0 + \Delta {\rm y} \right)^2 + \sigma^2_{\rm Y}
         - \left( X_0 + \Delta {\rm x} \right)^2  - \sigma^2_{\rm X}  \right],$
        }  \label{eq:intro_dpx}
   \end{eqnarray} 
   where we have integrated the Lorentz force over the transverse Gaussian bunch distribution of widths $\sigma_{\rm X}$ and  $\sigma_{\rm Y}$. The quantities $X_0$ and $Y_0$ denote the initial horizontal and vertical positions of the beam relative to the center of the sextupole prior to the strength change.
The sign of the horizontal orbit kick is given by the convention that it is positive toward the outside of the ring. Including only terms linear in $\Delta K_2L$, we have\\
   \begin{equation}
  \sigma^2_{\rm X} - \sigma^2_{\rm Y} = - 2\; \frac{\Delta  p_{\rm X}}{\Delta K_2L} + Y_0^2 - X_0^2. \label{eq:sigma}
  \end{equation}

   The generalization of this derivation to non-zero initial $K_2$ values is given in Sec.~\ref{sec:beamsize}.

   Our method for measuring beam sizes in sextupole magnets was inspired by a
private communication,\footnote{Reinhard Brinkman, private communication to Georg Hoffstaetter (2001)} and has not, to our knowledge, been developed elsewhere.

   Since early 2021, we have performed a set of measurements  of increasing sophistication and precision at the Cornell Electron-positron Storage Ring CESR,
   presenting the results in Refs.~\cite{Crittenden:IPAC21-MOPAB254}, \cite{Crittenden:IPAC22-MOPOTK040}, and~\cite{Crittenden:IPAC23-WEPL013}.
A web site describing the development of this project is available~\cite{chess-u-sextupole-studies}.
   Here we summarize our investigations into the contributions to the
   precision of our beam size calculations. The requirements of micron- and sub-microradian-level orbit measurement
   accuracy entail a detailed model of the CESR optics, including horizontal and vertical sextupole alignment values. Accurate determination of the sextupole calibration factors is important, as we have shown that the value obtained for the beam size is proportional to the value of $K_2$ (Eq.~\eqref{eq:sigma}).

   Throughout the analysis we employ a method for estimating measurement errors by 1)~observing the reproducibility of the measurements, and 2)~setting the residual weights in polynomial fits
   so as to obtain $\chi^2$/NDF=1~\cite{bevington_2003,nr_1987}.
In each case, we argue that the assumption that the uncertainties are independent of $K_2$ is reasonable in the relevant range.  Such scale factors are known as Birge factors~\cite{PhysRev.40.207}. Our fit parameters and associated uncertainty estimates are acquired using the MINUIT code~\cite{minuit,mnerror,JAMES1975343}.

We introduce CESR and the lattice optics in Sect.~\ref{sec:cesr}.

Section~\ref{sec:calfactors} covers the history of sextupole calibration values in use at CESR since the late 1990's, the measurement procedures in use since 2002, and the improved analysis methods introduced in 2022.

   Section~\ref{sec:misalignments} describes the measurement procedures used to obtain the sextupole alignment values and the data analysis techniques. The results are presented in Section~\ref{sec:misalignment_results}, together with the error estimates. Section~\ref{sec:misalignment_effects} discusses consequences for optics corrections of the misalignments at the observed level.

   Section~\ref{sec:beamsize} presents the full first-order derivation of the beam size, generalized to include $K_2$ changes where the initial value is not zero. Since the calculation of beam size requires the dependence of the
   angle change on the $K_2$ change, in addition to the quadrupole and skew quadrupole kicks, the means of determining the angle change is discussed, including error estimates. The results for the beam size values and uncertainties concludes this section.

   Finally, Sec.~\ref{sec:conclusions} presents discussions of the results and precision values obtained for the calibration factors, alignment determinations, and horizontal beam size values.
   
   \section{The Cornell Electron-positron Storage Ring}
   \label{sec:cesr}
The Cornell Electron-positron Storage Ring was commissioned in 1979 and ran at an energy of 5.289~\gev until June of 2001, at the $\Upsilon$ (4s) resonance, above the threshold of the production of bound states of bottom quarks. The period and its pre-history are chronicled in Ref.~\cite{berkelman}. In 2003, the storage ring operated at 1.9~\gev, collecting data on rare decays of bound states of charm quarks. CESR was converted in 2008 to a test accelerator (CESRTA)~\cite{Billing_2015} for the damping rings required by the proposed International Linear Collider. Following the termination of the CESRTA program, CESR resumed operation at 5.289~\gev as a X-ray light source using both the electron and positron beams. Following extensive modifications to the ring to convert to single-beam positron operations~\cite{PhysRevAccelBeams.22.021602}, the CHESS-U project began in 2019. One sixth of the ring was replaced, removing the straight section accommodating the detector studying electron-positron collisions (1979-2008) and improving the emittance by a factor of four. The investigations described here were performed during this era of CESR operation. The lattice parameters for CHESS-U operation are shown in Table~\ref{tab:lattice}.
\begin{table}[hbpt]
\centering
\caption{Lattice parameters for the CESR ring operating for CHESS-U.
        }
\label{tab:lattice}
\linethickness{3mm}
\renewcommand{\arraystretch}{1.1} \begin{tabular}{lc}
\hline
\hline
Beam energy (\gev) & 6.000\\
Circumference (m) & 768.44\\
Bunch current (mA)& 2.2 \\
Number of bunches & 45\\
Beam current (mA) & 100\\
RF frequency (MHz) & 500\\
Energy loss per turn (\mev) & 1.7 \\
Momentum compaction ($10^{-3}$) &  5.7\\
Bunch length (mm) & 17.1\\
Energy spread ($10^{-4}$) & 8.2 \\
Horizontal tune & 16.5557\\
Vertical tune & 12.6357\\
Synchrotron tune & 0.0336\\
Horizontal emittance (nm) & 28 \\
Vertical emittance (nm) & 0.1\\
\hline
\hline
\end{tabular}
\renewcommand{\arraystretch}{1.0}
\end{table}

The data recorded for the present study used a single bunch of  0.7~mA \mbox{(1.1 $\times 10^{10} e$)}, which avoided any saturation effects in the beam position monitor readout system. The turn-by-turn readout capability developed for CESRTA operations played an important role in our project.

A number of sextupole magnets served a dual purpose, including windings for either vertical steerings or skew quadrupoles. These contributions to the deflection of the positron beam were subtracted when fitting the dependence on sextupole strength change.

We adopt the numbering convention for the CESR sextupoles, which increases in the positron flight direction from the western end of the south arc (which has no sextupole magnets and comprises the undulators producing the X-ray beams), 9AW, 10W, 10AW, 12W, ..., to 47W in the north, continuing with 47E to 11E, 10AE, 9AE at the eastern beginning of the south arc. The sextupoles 9AW, 10W, 10AW, 10AE and 9AE are harmonic sextupoles, i.e. the dispersion is designed to be zero at these sextupoles. We also use the monotonic numerical scheme 8, 9, 10, ..., 88, 89, 91. We use the sextupole 10AW as the example in our analysis procedures.
   
   \section{Measurement of Sextupole Calibration Factors}
   \label{sec:calfactors}
   
   \subsection{Values in Use Prior to 2023}
   Prior to the calibration procedures described here, the values used were taken from transverse finite-element modeling and field measurements and 3D modeling for the field integrals~\cite{con-96-5, cbn-98-2}. Field uniformity was improved by redesigned pole faces in the late 1990's to accommodate operation with electron and positron beams sharing the 9-cm-wide beam pipe. The value used was $1.736 \times 10^{-4}$~m$^{-3}$/cu for a beam energy of 5.289~\gev, where cu are ``computer units,'' i.e. the digital command values issued to the power supply controllers. The controllers provided a maximum current of 12.5~A for 32k~cu. The 27.2-cm-long sextupoles have a field integral of 10.65~Tm at \mbox{$X = 1$~cm} at that current.
   
\subsection{Calibration Measurements 2002 - 2015}
The procedure developed for obtaining the calibration of each sextupole consisted of measuring the horizontal and vertical tune changes for a given change in $K_2$ for five beam positions set by a closed bump. This analysis neglects corrections arising from the beam motion consequential to the horizontal beam size.  These are typically less than 0.1~mm. A linear dependence of tune on the quadrupole error introduced by the sextupole strength change is also assumed, justified at our desired level of accuracy by running at a horizontal tune at least 25~kHz above the half-integer resonance at 195~kHz~\cite{crittenden_9sep2021, crittenden_14oct2021, crittenden_9feb2022}. The calibration correction factors were generally used to find polarity errors and shorted coils. They were not implemented in the CESR control system. During this period the sextupole calibration procedures were incorporated into the CESR modeling program CESRV based on the Bmad library~\cite{NIMA558:356to359} used during operations.

\subsection{Calibration Measurements 2022-2023 with Updated Analysis}
\label{sec:sextcalibrations}
A comprehensive re-calibration of the sextupoles was resumed in 2022 for the purposes of this beam-size measurement project. Two improvements were made:
\begin{itemize}
\item
A closed bump design specific to each sextupole was introduced, calculated on the fly during the tune-shift measurements. This required a model for the CESR optics which was obtained via the optimization procedure described in Sec.~\ref{sec:opt_ref}.
\item
The tune measurement and closed bump data were extracted from CESRV and subjected to a linear fit using the procedure described in Sec.~\ref{sec:introduction}.
\end{itemize}

\begin{figure}[htbp]
\centering
\includegraphics[width=\columnwidth]{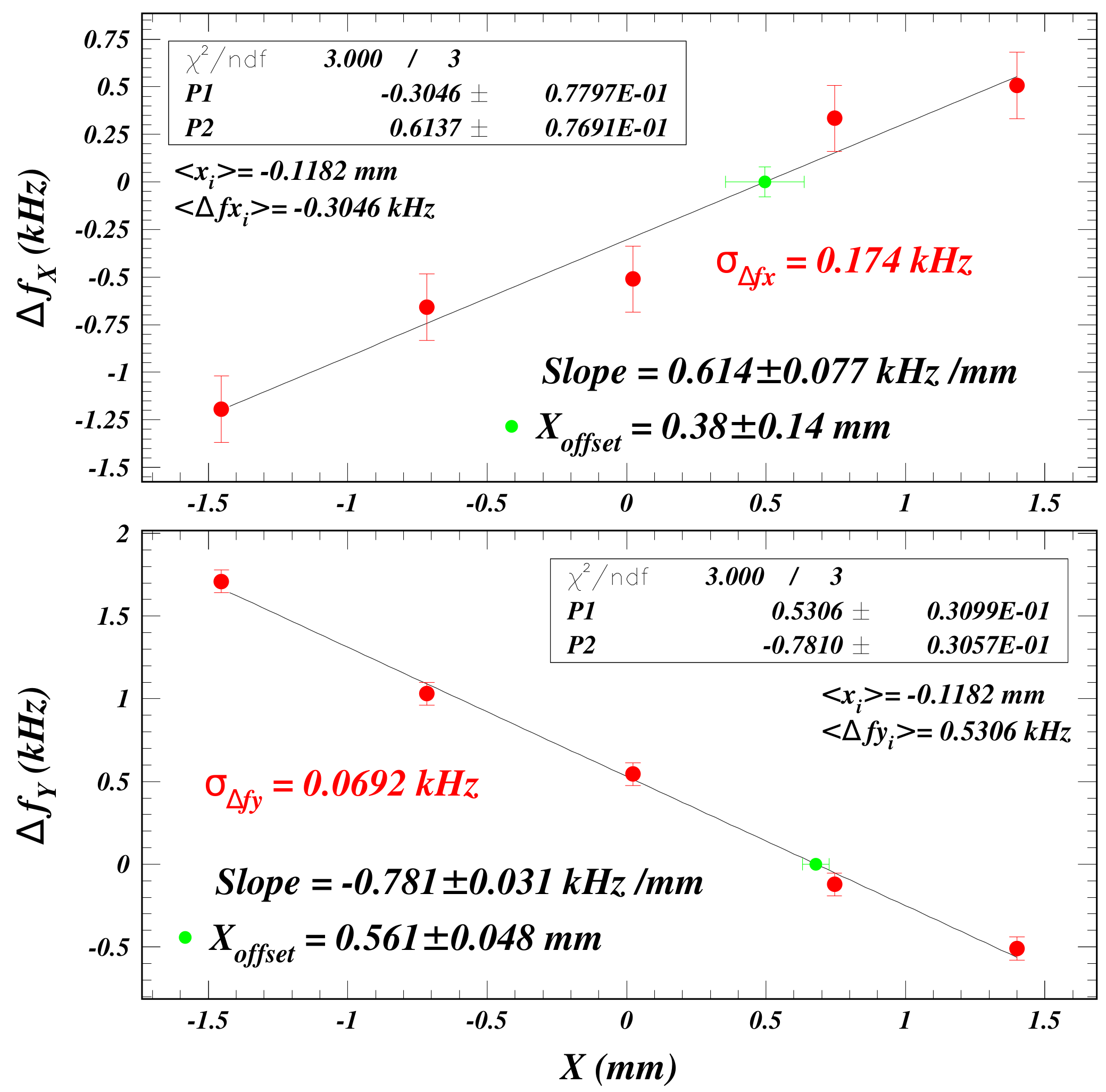}
\caption{Example of the measurement and analysis procedure to obtain the calibration correction factor. See text for details.
  }
   \label{fig:MOPOTK040_f4}
\end{figure}

The example of such measurements shown in
Fig.~\ref{fig:MOPOTK040_f4} 
uses the method of estimating uncertainties in the slope determinations by adjusting the residual weights. The slopes are of opposite sign and approximately in the ratio of the beta values, modulo a coupling contribution (e.g. vertical sextupole offset.) 

The calibration correction factors were obtained by using the beta-weighted difference of the horizontal and vertical tune shifts, which is insensitive to skew contributions and thus largely independent of vertical offset of the sextupole as can be inferred from the full 2D derivation of the tune shifts presented in the Sec.~\ref{sec:misalignments}.

Also shown in Fig.~\ref{fig:MOPOTK040_f4} as green points are the X~positions where the fit crosses zero, giving the horizontal sextupole offset in the BPM coordinate system. These were compared to the more accurate method discussed in Sec.~\ref{sec:misalignments} and found to have typical uncertainties of 0.1-0.3~mm. This method of reconstructing the beam position used only the two nearest BPMs, whereas the method in Sec.~\ref{sec:misalignments} used all available BPMs in the ring.

Horizontal and vertical tunes were measured by shaking the beam and locking to the tune~\cite{PAC11:MOP215}. Thirty-two single-pole-filtered 60-Hz samples were averaged, resulting in tune measurement RMS fluctuations between about 20 and 200~Hz. Those values are 174~Hz (horizontal) and 69~Hz (vertical) in the example shown in the previous section. This procedure required about 3~minutes per sextupole. The accuracy was shown to improve when additional 1-second averaging was included. For an additional 16 measurements, the tune accuracy improved by nearly a factor of four and the duration increased from three to ten minutes.

The calibration correction factor is derived from a measured/theory ratio for the beta-weighted tune shift differences~\cite{Crittenden:IPAC22-MOPOTK040}, where the theory value assumes the nominal calibration value used for the sextupoles during operations. A rough estimate of~5\% for the variations due to construction tolerances was made during the initial field measurements in 1998~\cite{cbn-98-2}.

\begin{figure}[htbp]
\centering
\includegraphics[width=\columnwidth]{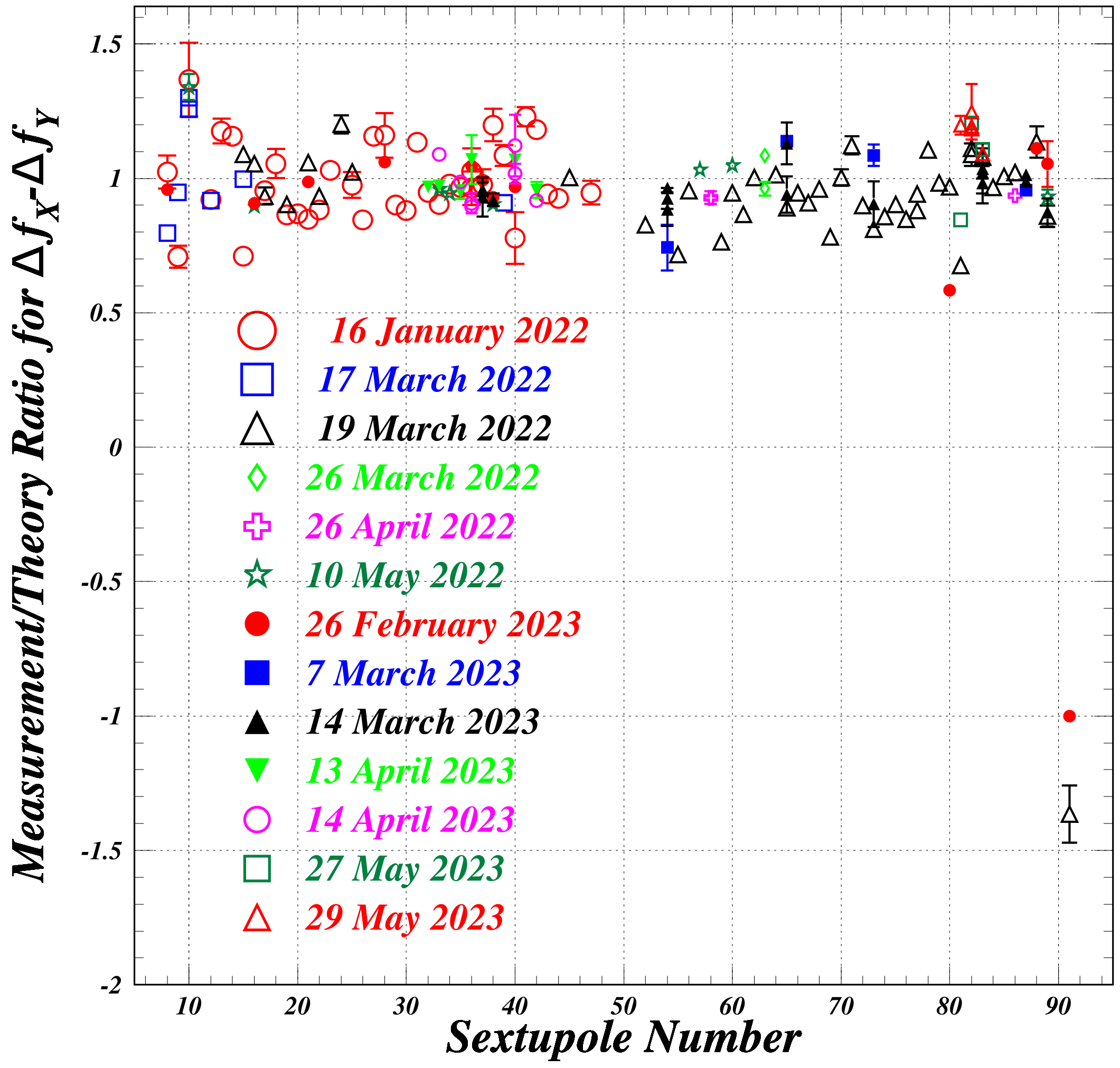}
\caption{Results for the calibration correction factor for each of the 76 CESR sextupoles. A total of 155~calibration data sets were recorded to measure repeatability and improve precision. The final values were determined using error-weighted averages.
  }
   \label{fig:sextcalfit_analysis_28}
\end{figure}

The measured calibration correction factors for the 76~sextupoles in the east and west arcs of the CESR ring  shown in Fig.~\ref{fig:sextcalfit_analysis_28}.
A total of 155~calibration data sets were recorded to measure repeatability and improve precision. The final values were determined using error-weighted averages.The assumed polarity for sextupole 91 was discovered to be incorrect.

\begin{figure}[htbp]
\centering
\includegraphics[width=\columnwidth]{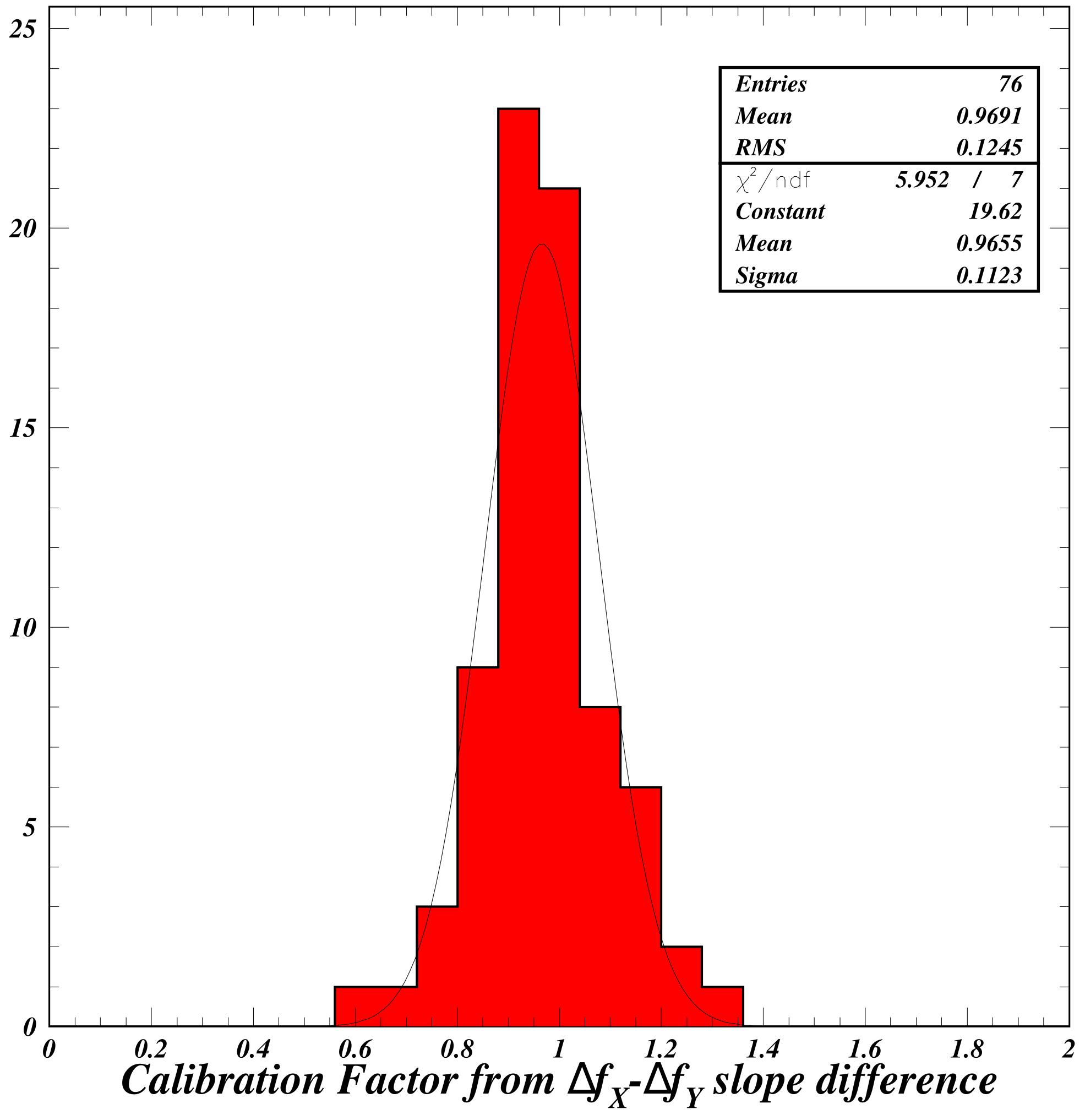}
\caption{Results for the calibration correction factor for each of the 76 CESR sextupoles. The RMS spread in correction factors is 12.5\% with an average value of 0.969.
  }
      \vspace*{-4mm}
   \label{fig:sextcalfit_analysis_46}
\end{figure}

Our measurements show an RMS deviation of~12.5\% with a mean value of~\mbox{$0.969$}, as observed in Fig.~\ref{fig:sextcalfit_analysis_46}, where the sign of the factor for sextupole 9AE (91) has been corrected.

\begin{figure}[htbp]
\centering
\includegraphics[width=\columnwidth]{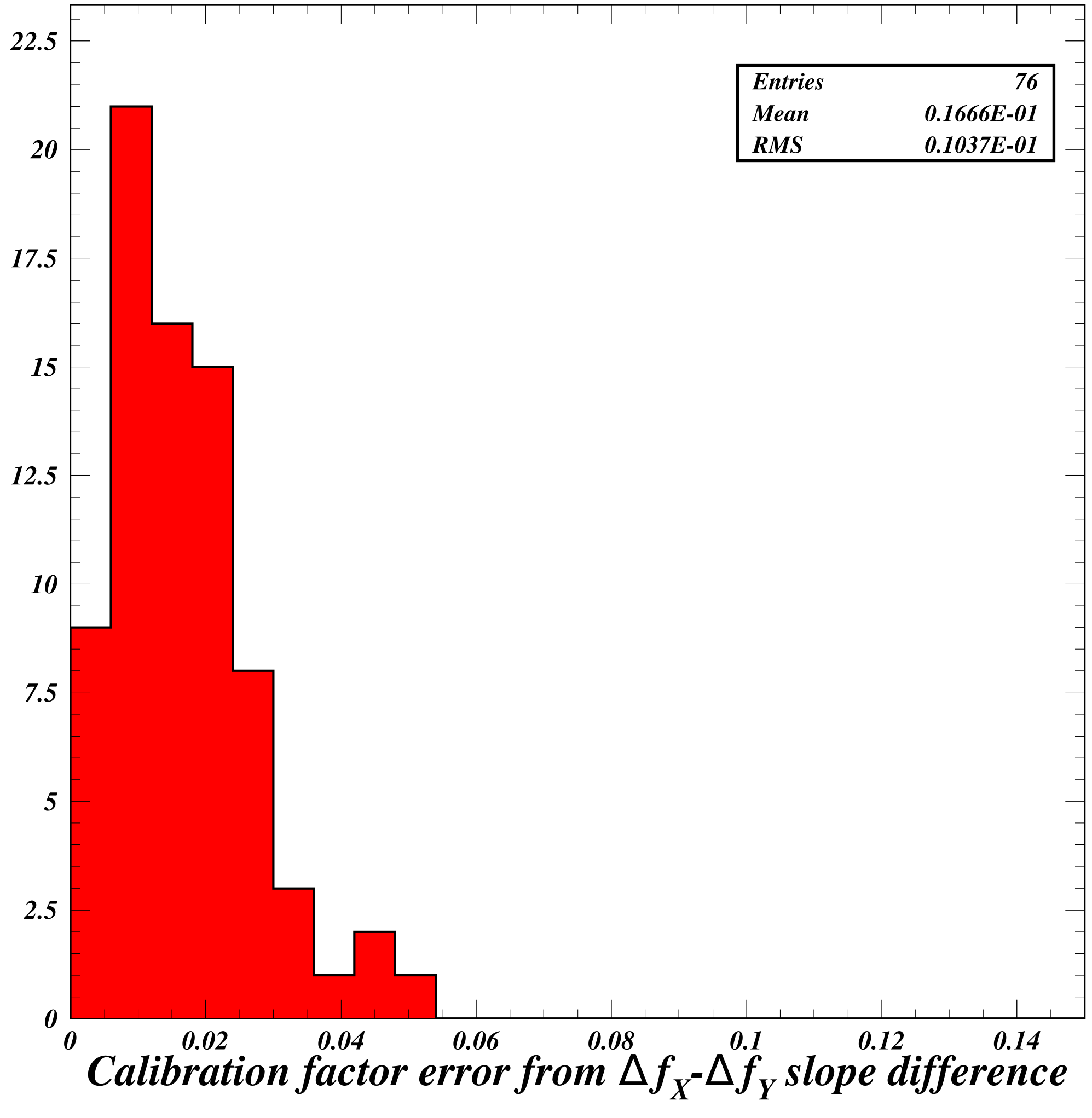}
\caption{Distribution in the uncertainties in the calibration correction factor measurements. The uncertainties average~1.7\% with an RMS spread of~1.0\%.
  }
      \vspace*{-4mm}
   \label{fig:sextcalfit_analysis_47}
\end{figure}

Figure~\ref{fig:sextcalfit_analysis_47}
shows the distribution in the uncertainty in the calibration correction factor determination.
The uncertainties average~1.7\% with an RMS spread of~1.0\%.

\section{Measurement of Sextupole Alignment Values}
\label{sec:misalignments}

\subsection{Introduction}
We calculate sextupole alignment values from the difference of the distance of the beam from the sextupole center, denoted $X_0$ and $Y_0$,
and the beam position at the longitudinal center of the sextupole in the reference system defined by the quadrupole centers, denoted $X_{\rm sext}$ and $Y_{\rm sext}$.
\begin{eqnarray}
  X_{\rm offset} &=& X_{\rm sext} - X_0 \\
  Y_{\rm offset} &=& Y_{\rm sext} - Y_0 
\end{eqnarray}

We begin with the first-order 2D analysis of betatron tune changes caused by the normal ($b_1$) and skew quad ($a_1$) terms arising from a sextupole strength ${\rm K}_2 L$~\cite{Crittenden:IPAC22-MOPOTK040}. We use the sextupole field components $\frac{ql}{p_0}B_{\rm x} =  K_2L xy$ and $\frac{ql}{p_0}B_{\rm y} = \frac{1}{2} K_2L (x^2 - y^2 )$
 and define the normal and skew quad multipole coefficients,
   \begin{equation}
     b_1 = \frac{qL}{P_0} \; \frac{\Delta B_{\rm Y}}{\Delta x} = {\rm K}_2  L \, x
   \end{equation}
   \begin{equation}
     a_1 = \frac{qL}{P_0} \; \frac{\Delta B_{\rm X}}{\Delta x} = {\rm K}_2 L \, y
   \end{equation}
we have the familiar results for the tune shifts from the normal quad term:
   \begin{equation}
     \Delta \mu_x = - \Delta b_1 \beta_x / 2 
   \end{equation}
   \begin{equation}
     \Delta \mu_y = \Delta b_1 \beta_y / 2 
   \end{equation}

The tune shifts from the skew quad terms can be shown~\cite{PhysRevSTAB.2.074001} to be 
   \begin{equation}
       \Delta \mu_x =  -(\Delta a_1)^2 \frac{\beta_x \beta_y \sin{ \mu_y}}{4\left(\cos{\mu_x}-\cos{\mu_y}\right)}
   \end{equation}
   \begin{equation}
       \Delta \mu_y =  (\Delta a_1)^2 \frac{\beta_x \beta_y \sin{ \mu_x}}{4\left(\cos{\mu_x}-\cos{\mu_y}\right)} 
   \end{equation}

Superposing the two contributions to the tunes and isolating $a_1$ and $b_1$, we obtain their values as functions of known quantities when the tune shifts are measured:
  \begin{equation}
    \sin{\mu_x} \Delta \mu_x + \sin{\mu_y} \Delta \mu_y =  \frac{- \Delta b_1}{2} \left(\beta_x \sin{\mu_x}   - \beta_y \sin{\mu_y}  \right)
   \end{equation}
   \begin{equation}
    \beta_y \Delta \mu_x + \beta_x \Delta \mu_y = (\Delta a_1)^2 \; \frac{\beta_x \beta_y \left( \beta_x \sin{ \mu_x} -  \beta_y \sin{ \mu_y} \right)}{4\left(\cos{\mu_x}-\cos{\mu_y}\right)} 
   \end{equation}

   The second equation shows that
   \begin{equation}
     \label{eq:ipac22_db1}
     \Delta b_1 = \frac{\Delta \mu_y}{\beta_y} - \frac{\Delta \mu_x}{\beta_x}
   \end{equation}
   is more independent of $a_1$ than $b_1$ derived from either $\Delta \mu_x$ or $\Delta \mu_y$ alone. 

The 2D calculations of $b_1$ and $a_1$ are sensitive to cancellation divergences.
The values of the initial tunes are such that $\sin{\mu_x} \simeq -0.4$ and $\sin{\mu_y} \simeq -0.8$. Thus the formula for $b_1$ and $a_1$ both diverge for $\beta_x \simeq 2 \beta_y$.

For simplicity of presentation, we have used here the approximation
\begin{equation}
  \cos{(\mu + \Delta \mu)} - \cos{\mu} \simeq \Delta \mu \sin{\mu}.
\end{equation}
This approximation breaks down near the half-integer resonance. In fact,
the quadratic term in Fig.~1 of Ref.~\cite{Crittenden:IPAC21-MOPAB254} was later shown to arise from this approximation,
rather than from the quadratic term $\Delta K_2L \; \Delta x$. With this approximation removed, the quadratic term is consistent with the horizontal beam motion arising from the sextupole strength term. Our choice, however, is to use only linear terms in the calculation of beam size, since these are much more accurately determined.

\subsection{Modeling Studies for Sextupole Alignment Analysis}
\label{sec:model_alignment}
\begin{figure}[htbp]
\centering
\includegraphics[width=\columnwidth]{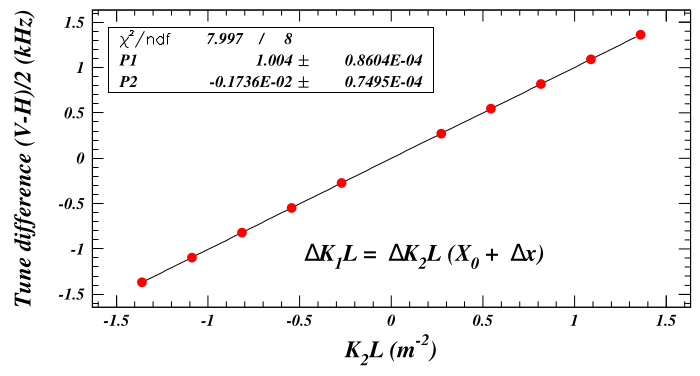}
\caption{The quadrupole term $\Delta K_1 L$ calculated from the tune shifts as a function of the sextupole 10AW strength change modeled by tracking 2000 beam particles through the CESR design lattice. The simulated 1-mm offset on the sextupole is reconstructed with an inaccuracy of~4~microns.  The uncertainties given the for the coefficients are due to machine accuracy in the modeling.
  }
   \label{fig:ana_k2scan_10_10aw_13}
\end{figure}

Nonlinear effects, such as beam motion in sextupoles arising from strength change in the studied sextupole, perturb the linear analysis presented below in Sec.~\ref{sec:analysis_quad}. Modeling studies are necessary to quantify the magnitudes of such effects.
A model was developed~\cite{crittenden_11apr2023}, tracking 2000 beam positrons through the CESR design lattice. A 1~mm horizontal misalignment in CESR sextupole 10AW was put in the model.
Choosing eleven $K_2$ settings, we write out tunes and sextupole attributes including beam coordinate centroid and RMS values at the sextupole.
To reconstruct $X_0$, we use the method of subtracting beta-weighted horizontal and vertical tune changes to remove coupling contributions (see Sec.~\ref{sec:quad_fit} and Eqs.~\ref{eq:ipac22_db1} and~\ref{eq:tunediff}).
Figure~\ref{fig:ana_k2scan_10_10aw_13}
shows the value of the quadrupole term $\Delta K_1 L$ as a function of the sextupole strength change.
This model also shows that our approximation using the beta-weighted tune difference is sufficiently accurate. Since the beam is on-axis, aside from the small beam motion in the sextupole, we recover the 1-mm offset with an inaccuracy of~4~microns.

\subsection{Measurement Procedure}
\label{sec:measurement_procedure}
For each $K_2$ setting, multiple measurements of phase, orbit and coupling functions are recorded~\cite{PhysRevSTAB.3.092801}. At least three such measurements are taken in all cases. Each of these measurements is highly averaged in the front-end data acquisition algorithm. Studies of the uncertainty dependence on the number of $K_2$ setttings have also been performed. It was found, for example, that five $K_2$ settings are insufficient to determine accurately the quadratic coefficients, which require nine $K_2$ settings to achieve a precision of 10\% or better.

\subsection{Optimizations for Reference Functions}
\label{sec:opt_ref}
The model for the optics reference functions ($\Delta K_2 = 0$~m$^{-3}$) was obtained by varying the steering, quadrupole and skew quadrupole magnet settings in the model to best match the measured phase, orbit and coupling measurements. Prior to the optimization, geometrical misalignments obtained from periodic laser alignment measurements of dipole magnet roll values were included in the model. The horizontal and vertical sextupole magnet offsets measured as described below in Sec.~\ref{sec:ana_diff} were also included. Finally, it was found that loading the sextupole magnet settings recorded in the measurement file improved the ultimate merit function achieved. Since these values are recorded in computer units, the calibration values described in Sec.~\ref{sec:sextcalibrations} were required.

Each BPM contributes five constraints contributing to the merit function: horizontal and vertical phase, horizontal and vertical orbit, and coupling.
There are typically 295 variables and 380 constraints in the final iteration of the optimization. The iterative procedure consisted of rejecting anomalous contributions to the merit function, then rerunning the optimization.

Since phase, coupling, orbits and tunes are fit simultaneously, the relative weights for constraints are important. These were obtained using the optimization for difference functions, as described in Secs.~\ref{sec:analysis_quad} and~\ref{sec:analysis_orbit}. The weights used correspond to precision values of 0.1~mm for the orbits, 0.02\degree for the phase functions, 0.02 for the coupling $\bar{C}_{12}$, and 0.001 for the tune $Q$. The precision in the coupling measurements was assumed to be the same as for the phase function, since the residual analysis for the quadrupole and skew quadrupole terms obtained similar values (see Sec.~\ref{sec:ana_diff}). In the case of the orbit precision, the precision values of a few microns were found in the fits to the difference functions~\ref{sec:analysis_orbit}. The weights for the optimizations for the reference functions were increased to 0.1~mm to account for systematics in the determination of the BPM offsets relative to the quadrupole centers and for contributions from the button gain measurements. The latter were found to depend on the thermal state of the machine, varying over 12~hours after the begin of full-current operation. The sextupole scan data was obtained in the cold state of the machine.

The means of estimating appropriate weights for the contributions of phase, orbit and coupling functions to the merit functions are discussed below.

\subsubsection{Orbit Residuals Analysis}
\begin{figure}[htbp]
\centering
\includegraphics[width=\columnwidth]{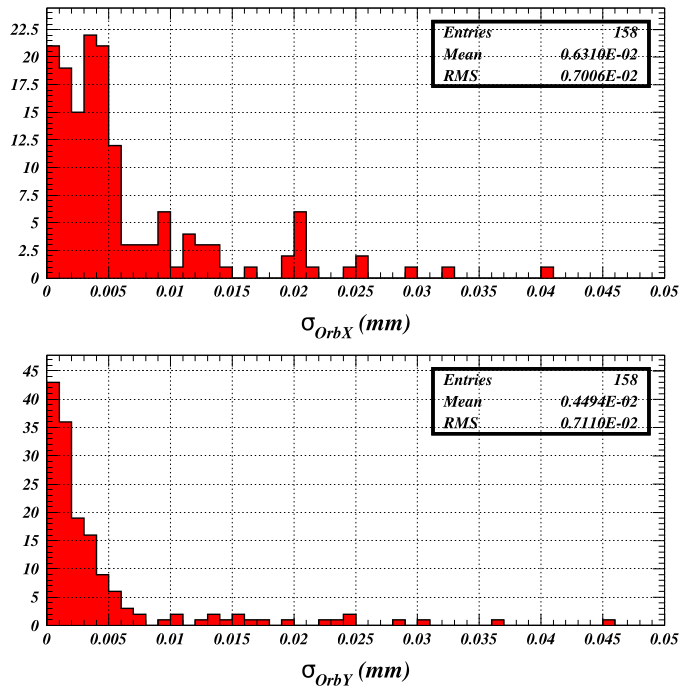}
\caption{Distributions in the precision in the determination of the horizontal ($\sigma_{\rm OrbX}$) and vertical ($\sigma_{\rm OrbY}$) orbit changes caused by the sextupole strength change $\Delta K_2$ for all scans. 
  }
   \label{fig:anamain_1_191_5sep2024_6}
\end{figure}

\begin{figure}[htbp]
\centering
\includegraphics[width=\columnwidth]{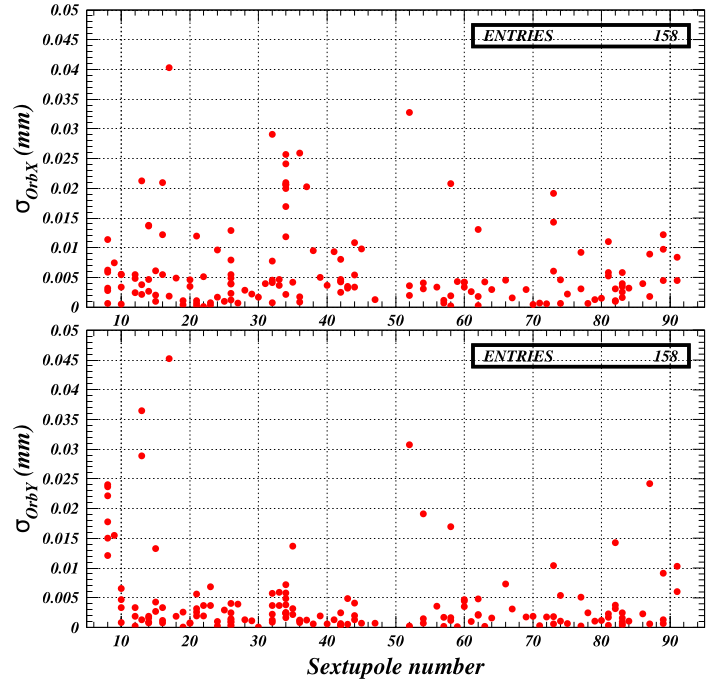}
\caption{The uncertainties in the determination of the horizontal and vertical orbit changes caused by the sextupole strength change $\Delta K_2$ versus sextupole number.
  }
   \label{fig:anamain_1_191_5sep2024_7}
\end{figure}

\begin{figure}[htbp]
\centering
\includegraphics[width=\columnwidth]{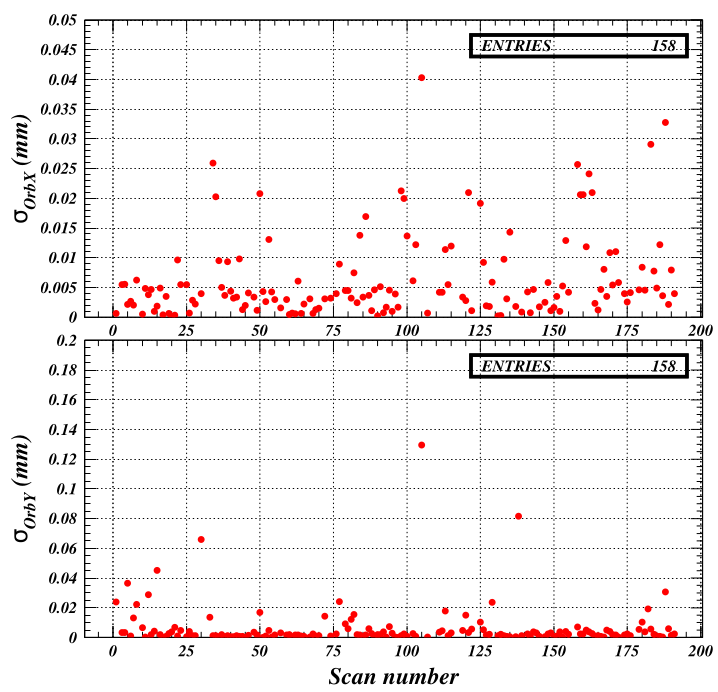}
\caption{The uncertainties in the determination of the horizontal and vertical orbit changes caused by the sextupole strength change $\Delta K_2$ versus scan number.
  }
   \label{fig:anamain_1_191_5sep2024_8}
\end{figure}

The example of the $K_2$ scan result for orbit dependence on the 10AW sextupole strength change are shown below in Figs.~\ref{fig:main_85_40} and~\ref{fig:main_85_42}
in Sec.~\ref{sec:quad_fit}, where the contributions to the uncertainties in the misalignment value determinations are discussed. The RMS variation in the average of three measurements was found to be 3.3~$\mu$m (0.8~$\mu$m) for the horizontal (vertical) orbit measurement at each $K_2$ setting.
For the results for all sextupoles and all scans, see Figs.~\ref{fig:anamain_1_191_5sep2024_6},~\ref{fig:anamain_1_191_5sep2024_7}, and~\ref{fig:anamain_1_191_5sep2024_8}.
Typical values for the uncertainty in the horizontal (vertical) position is found to be less than 6~$\mu$m (4~$\mu$m). The tail results in the distribution RMS value of 7~$\mu$m (7~$\mu$m). However, due to a variety of systematic effects in the BPM data acquisition mentioned in Sec.~\ref{sec:opt_ref}, a value of 0.1~mm was used for the weights in the calculation of the merit function.

\subsubsection{Phase Function Residuals Analysis}
\begin{figure}[htbp]
\centering
\includegraphics[width=\columnwidth]{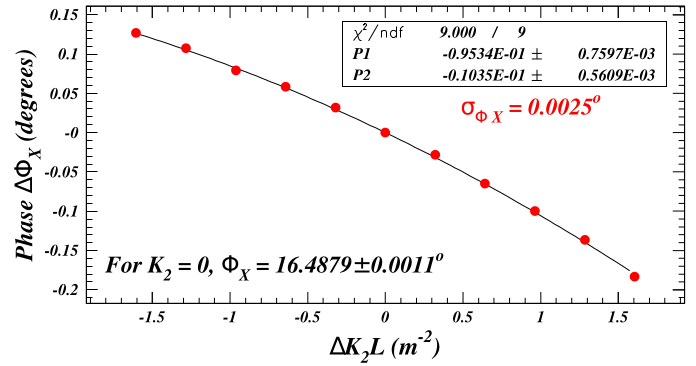}
\caption{The change in the horizontal phase at sextupole 10AW induced by the change in sextupole strength $\Delta K_2$. Adjusting the assumed uncertainty in each point to give $\chi^2$/NDF to unity results in a value for the RMS variation in the average of three measurements of 0.0025~degrees.
  }
   \label{fig:main_85_44}
\end{figure}

\begin{figure}[htbp]
\centering
\includegraphics[width=\columnwidth]{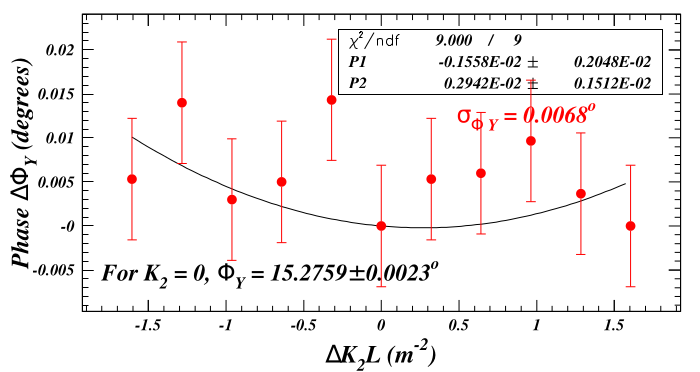}
\caption{The change in the vertical phase at sextupole 10AW induced by the change in sextupole strength $\Delta K_2$. Adjusting the assumed uncertainty in each to give $\chi^2$/NDF to unity results in a value for the RMS variation in the average of three measurements of~0.0068~degrees.
  }
   \label{fig:main_85_46}
\end{figure}

\begin{figure}[htbp]
\centering
\includegraphics[width=\columnwidth]{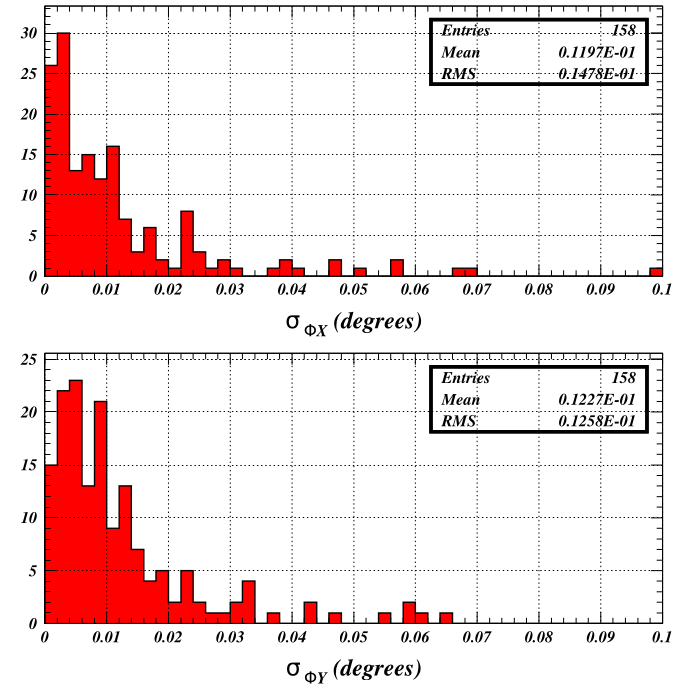}
\caption{Distributions in the precision in the determination of the horizontal ($\sigma_{\rm \Phi{X}}$) and vertical ($\sigma_{\rm \Phi{Y}}$) phase changes caused by the sextupole strength change $\Delta K_2$ for all scans. 
  }
   \label{fig:anamain_1_191_5sep2024_3}
\end{figure}

\begin{figure}[htbp]
\centering
\includegraphics[width=\columnwidth]{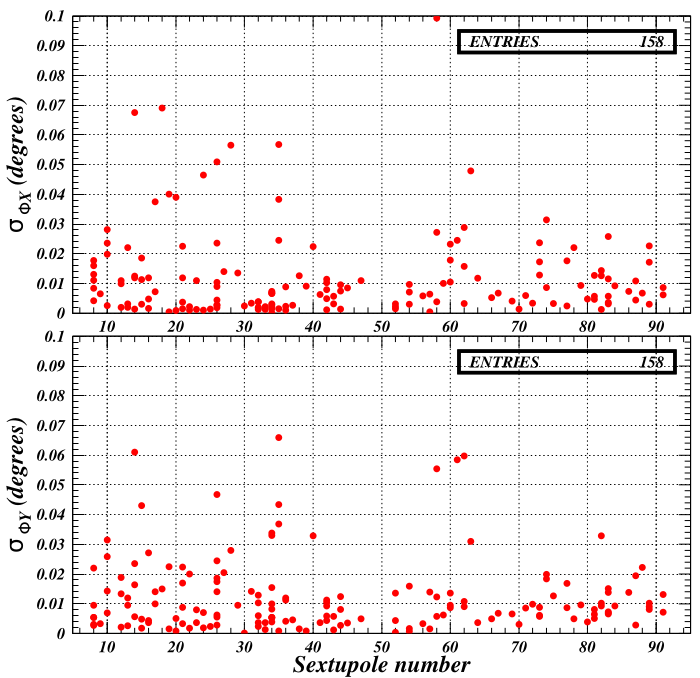}
\caption{The uncertainties in the determination of the horizontal and vertical phase function changes caused by the sextupole strength change $\Delta K_2$ versus sextupole number.
  }
   \label{fig:anamain_1_191_5sep2024_4}
\end{figure}

\begin{figure}[htbp]
\centering
\includegraphics[width=\columnwidth]{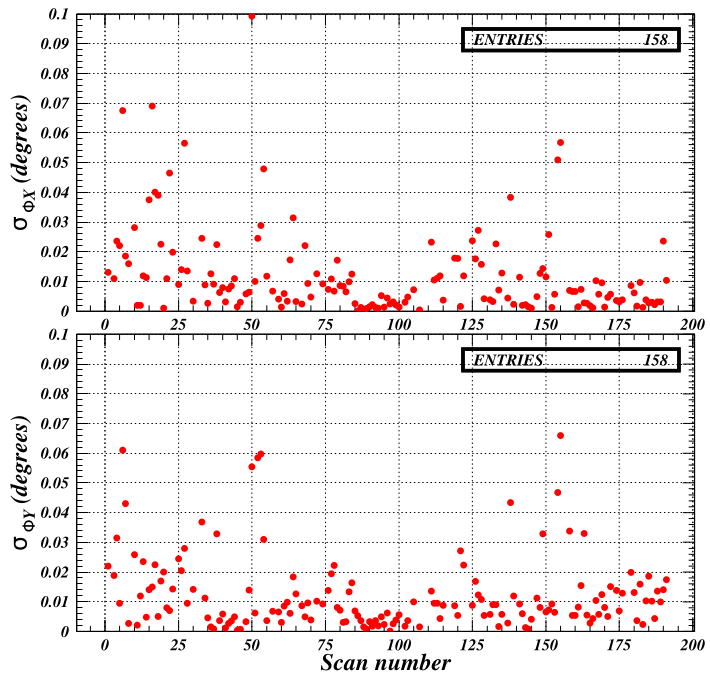}
\caption{The uncertainties in the determination of the horizontal and vertical phase function changes caused by the sextupole strength change $\Delta K_2$ versus scan number.
  }
   \label{fig:anamain_1_191_5sep2024_5}
\end{figure}

Examples of the phase function change $\Delta \Phi_X$ as a function of the sextupole strength change $\Delta K_2$ from the 10AW scan 85 are shown in Figs.~\ref{fig:main_85_44} and~\ref{fig:main_85_46}.
The RMS precision in the average of three measurements was found to be 0.0025\degree (0.0068\degree) for the horizontal (vertical) phase measurement at each $K_2$ setting. These values are smaller than typical. For the results for all sextupoles and all scans, see Figs.~\ref{fig:anamain_1_191_5sep2024_3},~\ref{fig:anamain_1_191_5sep2024_4}, and~\ref{fig:anamain_1_191_5sep2024_5}.
Typical values for the uncertainty in the horizontal (vertical) phase change are found to be less than 0.014\degree (0.016\degree). The tail results in the distribution RMS value of 15\degree (13\degree).

The observed level of repeatability is about 0.01\degree for the average of three difference measurements. Ref.~\cite{PhysRevSTAB.3.092801} cites a contribution from the betatron clock noise of  approximately 0.006\degree. Many improvements have been introduced in the BPM data acquisition system since that time. The weight assumed for the contribution to the merit function in the fit to the reference phase function is 0.02\degree.

\subsection{Optimizations for Difference Functions}
\label{sec:opt_diff}
The optimizations for the difference functions use four variables: quadrupole, skew quadrupole, and and horizontal and vertical dipole terms superposed on the sextupole, caused by the change in sextupole strength $\Delta K_2$. The typical number of constraints for these fit was about 380 phase, coupling and orbit measurements.

The precision in the quadrupole term $\Delta b_1$ and the skew quadrupole term $\Delta a_1$ are discussed in Sects.~\ref{sec:quad_fit} and~\ref{sec:skew_quad_fit} on the measurement of sextupole misalignments. The horizontal and vertical dipole terms resulting from the sextupole strength change are not used directly in the calculation of the beam size, however, together with the quadrupole and skew quadrupole terms, they give the orbit angle changes in the sextupole which are used in the beam size determination.

\subsection{Analysis Procedure for Difference Functions}
\label{sec:ana_diff}
The orbit, phase, and coupling difference functions were obtained from three repetitive measurements at each sextupole strength setting. The averages of the three were then subjected to the polynomial fit procedure described below.
\subsubsection{Scatter Plots}
The value of the scatter plots is two-fold: 1)~an order-of-magnitude estimate of
the precision of the measurements can be gleaned from the observed repeatability, and 2)~the occasional anomalous failed optimization is easily identified. Our more accurate method of determining measurement precision is presented below in Sec.~\ref{sec:analysis_quad}.

\begin{figure}[htbp]
\centering
\includegraphics[width=\columnwidth]{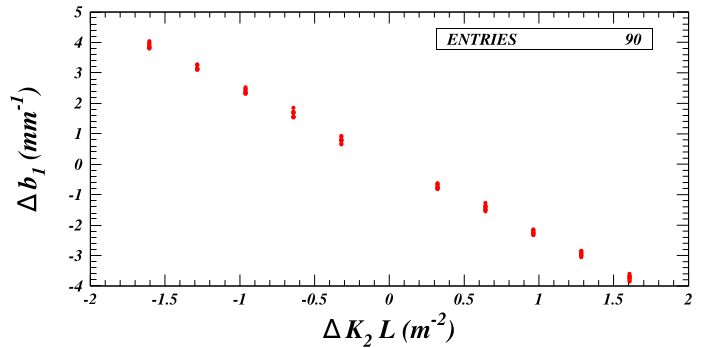}
\caption{Scatter plots for the three difference measurements at eleven $K_2$ values for the quadrupole term $\Delta b_1$ induced by the change in sextupole strength $\Delta K_2$. The repeatability is observed to be better than about~0.4~mm$^{-1}$.
  }
   \label{fig:main_85_15_db1}
\end{figure}

\begin{figure}[htbp]
\centering
\includegraphics[width=\columnwidth]{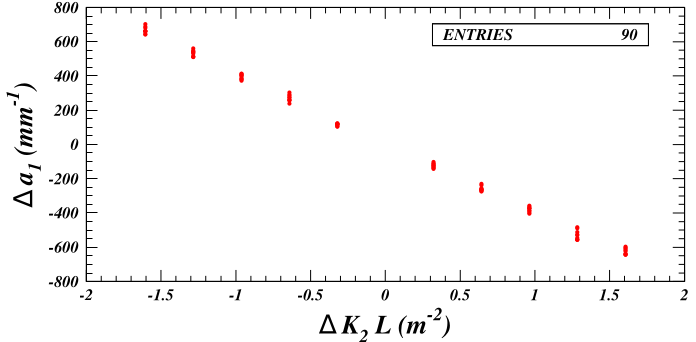}
\caption{Scatter plots for the three difference measurements at eleven $K_2$ values for the quadrupole term $\Delta a_1$ induced by the change in sextupole strength $\Delta K_2$. The repeatability is observed to be better than about~0.2~mm$^{-1}$.
  }
   \label{fig:main_85_15_da1}
\end{figure}

Figures~\ref{fig:main_85_15_db1} and~\ref{fig:main_85_15_da1}
show scatter plots in $\Delta b_1$ and $\Delta a_1$ for the example of scan 85, recorded for sextupole 10AW on the 23rd of October, 2021. Three phase, orbit and coupling measurements were made at each of eleven $K_2$ settings. The observed repeatability of better than 0.4~mm$^{-1}$ suffices for our desired precision, as shown below in Sec.~\ref{sec:analysis_quad}.
  
\subsubsection{Polynomial Fits and Error Analysis}
\label{sec:analysis_quad}
The sections~\ref{sec:quad_fit} and~\ref{sec:skew_quad_fit} below describe the polynomial fits to the dependence of quadrupole and skew quadrupole terms arising from changes in sextupole strength. The precision of our measurements allows very good determination of the linear term and often good determination of the quadratic coefficient. At this level of precision, no significant cubic term is observed, as expected from the linear analysis (Eqs.~\ref{eq:db1} and~\ref{eq:da1}).
This observation encourages the approximation that nonlinear effects such as those mentioned in~\ref{sec:model_alignment} are small.

\subsubsection{Quadrupole Term}
\label{sec:quad_fit}
\paragraph{Tune Measurements}

Our first method of determining $X_0$, the horizontal distance of the beam
from the center of the sextupole prior to changing the strength of the sextupole $K_2$,
is to derive the $\Delta K_1$ value from the beta-weighted difference of horizontal and vertical  tune measurements according to
\begin{equation}
  \label{eq:tunediff}
     \Delta K_1 L = \frac{\Delta \mu_y}{\beta_y} - \frac{\Delta \mu_x}{\beta_x}
   \end{equation}
   derived in Ref.~\cite{Crittenden:IPAC22-MOPOTK040}.  This calculation is more insensitive to skew quadrupole contributions than the value derived from either $\Delta \mu_x$ or $\Delta \mu_y$ alone. 

\begin{figure}[htbp]
\centering
\includegraphics[width=\columnwidth]{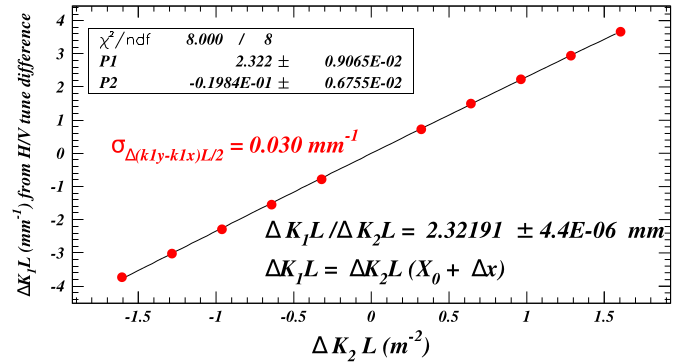}\\
\caption{Quadrupole kick values $K_1$ derived from betatron tune changes as a function of sextupole strength change for the example of scan 85. 
  }
   \label{fig:main_85_9}
\end{figure}

Our tune measurements derive from two sources: 1)~we operate the Digital Tune Tracker~\cite{PAC11:MOP215} continuously during the measurements, obtaining about 20~measurements at intervals of 3~seconds for each sextupole setting, 2)~following three phase function measurements at each sextupole setting, we record turn-by-turn orbit data, 32k~revolutions for each of 126~beam position monitors (BPMs).  This data is post-processed to obtain tune measurements with an accuracy of about one part in 10$^{4}$. The combination of these two tune measurement methods provides an accuracy of about 0.003\%. Figure~\ref{fig:main_85_9}
shows an example of ten difference measurements obtained from eleven sextupole settings. We employ a method for estimating uncertainties in the polynomial coefficients by adjusting the residual weights to obtain $\chi^2$/NDF=1. The linear term provides us with a value for $X_0$ of \mbox{$-2.3219 \pm 0.0091$~mm.} The estimate for the $\Delta K_1 L$ uncertainty in each point is 0.030~mm$^{-1}$.

\paragraph{Difference Phase Functions}

A second, independent, means of determining $X_0$ is to record phase function and orbit measurements at each sextupole setting, then to fit the difference functions with multipole values $b_1$, $a_1$ and horizontal and vertical dipole kicks superposed on the sextupole.
We choose to distinguish these two methods by denoting the kick $\Delta K_1 L$ for the case of the tune change and $\Delta b_1$ for the case of the difference phase function. We show below that the two methods provide consistent results with $\lsim 0.1$~mm precision.

\begin{figure}[htbp]
\centering
\includegraphics[width=\columnwidth]{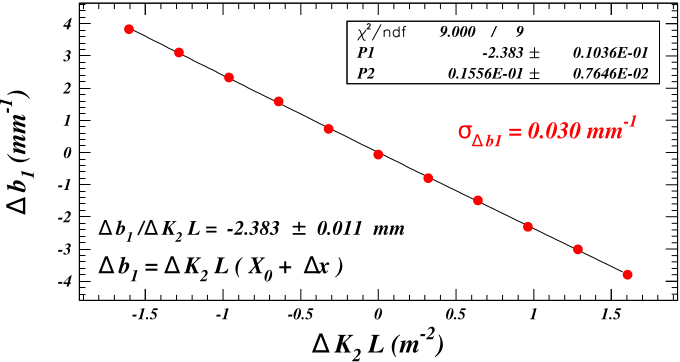}
\caption{Example of scan 85 for the polynomial fit results for the quadrupole term $\Delta b_1$ induced by the change in sextupole strength $\Delta K_2$. Quadrupole kick values $\Delta b_1$ at the sextupole caused by a change in sextupole strength $\Delta K_2$. These are determined using a fit to phase function, coupling function and orbit differences while varying $\Delta b_1$, a skew quadrupole kick $\Delta a_1$, and horizontal and vertical dipole kicks.
  }
   \label{fig:main_85_19}
\end{figure}

Figure~\ref{fig:main_85_19}
 shows the results for the quadrupole term $\Delta b_1$ obtained from the fit to the difference phase, orbit and coupling functions. The polynomial fit procedure described in Sec.~\ref{sec:introduction} provides a value for $X_0$ of \mbox{$-2.383 \pm 0.010$~mm}.

\begin{figure}[htbp]
\centering
\includegraphics[width=\columnwidth]{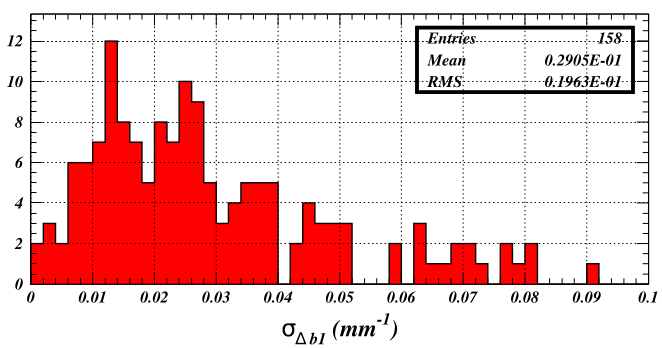}
\caption{Distribution in precision values for the quadrupole term change $\sigma_{\Delta b1}$ for all scans.
  }
   \label{fig:anamain_1_191_5sep2024_9_db1}
\end{figure}

\begin{figure}[htbp]
\centering
\includegraphics[width=\columnwidth]{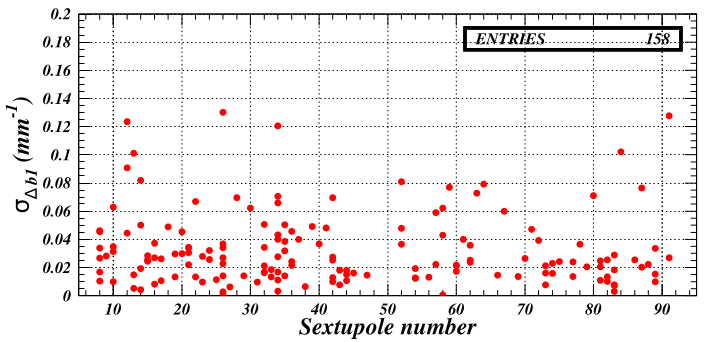}
\caption{Precision values for the quadrupole term change $\sigma_{\Delta b1}$ for each sextupole.
  }
   \label{fig:anamain_1_191_5sep2024_11_db1}
\end{figure}

\begin{figure}[htbp]
\centering
\includegraphics[width=\columnwidth]{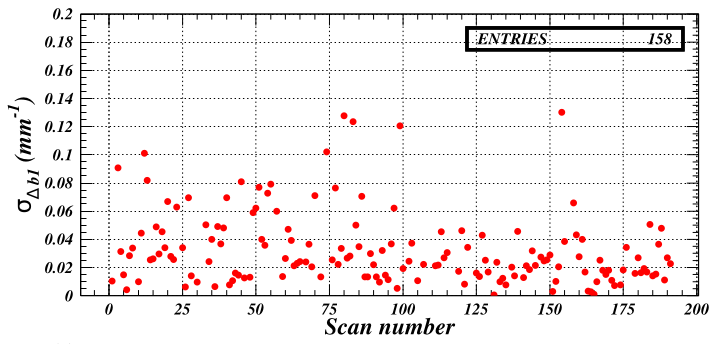}
\caption{Precision values for the quadrupole term change $\sigma_{\Delta b1}$ for each $K_2$ scan.
  }
   \label{fig:anamain_1_191_5sep2024_13_db1}
\end{figure}

The precision of each $\Delta b_1$ point obtained from setting $\chi^2$/NDF=1 for this scan is 0.030~mm$^{-1}$. This value is typical of the scans. Figure~\ref{fig:anamain_1_191_5sep2024_9_db1}
shows the distribution in precision values for all scans. The average value is 0.029~mm$^{-1}$ and the RMS is 0.019~mm$^{-1}$. The values obtained from all scans are shown for each sextupole in Fig.~\ref{fig:anamain_1_191_5sep2024_11_db1}.
The values obtained from all scans are shown as a function of scan number in Fig.~\ref{fig:anamain_1_191_5sep2024_13_db1}.

\begin{figure}[htbp]
\centering
\includegraphics[width=\columnwidth]{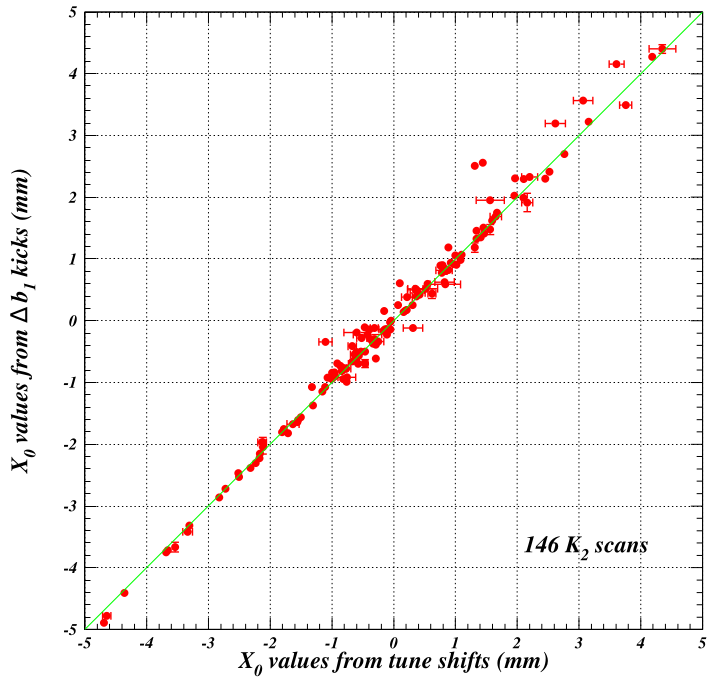}
\caption{Degree of correlation obtained from the values for $X_0$ derived from tune changes and from fits to phase function, orbit, and coupling differences for the change in the quadrupole term $\Delta b_1$.
  }
   \label{fig:anamain_1_191_5sep2024_26}
\end{figure}

\begin{figure}[htbp]
\centering
\includegraphics[width=\columnwidth]{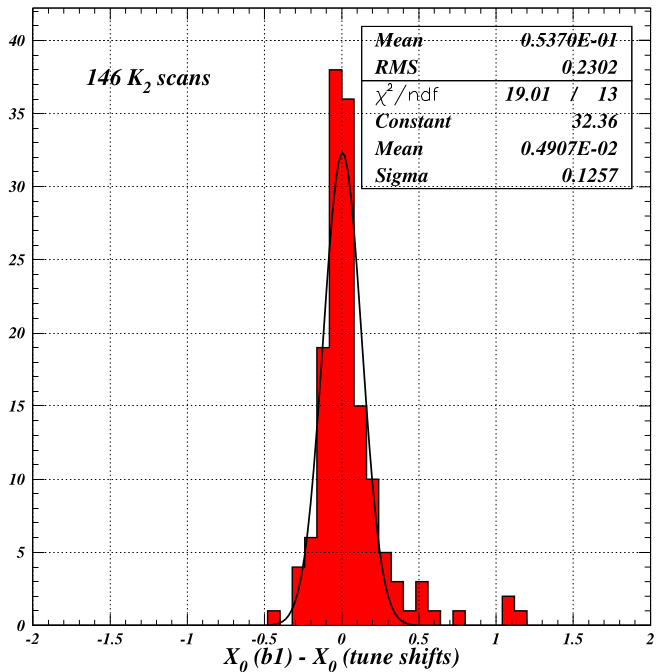}\\
\caption{Difference distribution for the $X_0$ values determined from the $K_1 L$ values from the tune shifts and the $\Delta b_1$ values measured using the difference functions. These measurements were found to be uncorrelated by comparing their uncertain values, so the RMS value allows the determination of an upper bound on the precision of the $X_0$ measurements.
  }
   \label{fig:anamain_1_191_5sep2024_28}
\end{figure}

These two methods for determining $X_0$ are compared in the correlation plot in Fig.~\ref{fig:anamain_1_191_5sep2024_26}
which includes all measurements to date.
The RMS of the difference distribution, shown in Fig.~\ref{fig:anamain_1_191_5sep2024_28}, is 0.11~mm (excluding anomalies),
showing sufficient precision for measuring beam sizes of 1-2-mm.
The mean uncertainty for the  $X_0$ values determined from the $K_1 L$ values from the tune shifts is 0.044~mm.The mean uncertainty for the  $X_0$ values determined from  the $\Delta b_1$ values is 0.015~mm. The significance of the good agreement between the local kick result and the ring-wide tune measurement is that the underlying assumption of linear optics is sufficiently accurate for our purposes.

The horizontal misalignment $X_{\rm offset}$ of the sextupole relative to the BPM coordinate system, which defines the origin as the centers of the quadrupole magnets, can now be found by determining the horizontal orbit position measurement prior to the sextupole strength change. The full statistical power of the measurements at eleven sextupole settings is shown in Fig.~\ref{fig:main_85_40}

\subsubsection{Skew Quadrupole Term from Difference Coupling Functions}
\label{sec:skew_quad_fit}
\begin{figure}[htbp]
\centering
\includegraphics[width=\columnwidth]{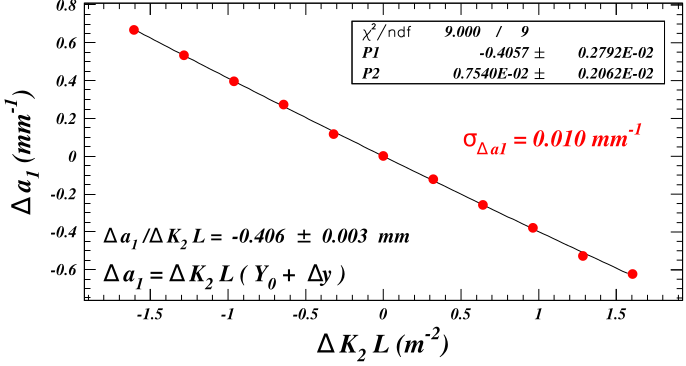}
\caption{Example of a polynomial fit result for the change in the skew quadrupole term $\Delta a_1$ 
  induced by the change in sextupole strength $\Delta K_2$. This example is for the $K_2$ scan 85 of sextupole 10AW. The vertical position of the beam relative to the center of the sextupole is measured to be \mbox{$Y_0 = -0.406 \pm 0.003$~mm}.
  }
   \label{fig:main_85_23}
\end{figure}

\begin{figure}[htbp]
\centering
\includegraphics[width=\columnwidth]{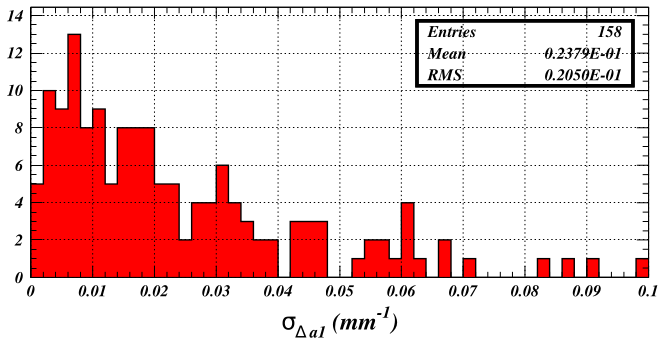}
\caption{Distribution in precision values for the skew quadrupole term change $\Delta a_1$ for all scans.
  }
   \label{fig:anamain_1_191_5sep2024_9_da1}
\end{figure}

Figure~\ref{fig:main_85_23}
shows the dependence of the skew quadrupole kick dependence on the sextupole strength in sextupole 10AW measured during scan 85. The value of $Y_0$ is determined with a precision of 3~$\mu$m. The statistical uncertainty in the average of three measurements of $\Delta a_1$ is found to be \mbox{0.010~mm$^{-1}$}. This value is typical, as seen in Fig.~\ref{fig:anamain_1_191_5sep2024_9_da1},
which shows the result for all 158~scans. The average value is 24~$\mu$m; the RMS of the distribution is 21~$\mu$m. Such values are sufficiently precise for the measurement of beam size and generally smaller than the contribution of the measurement precision of the horizontal angle change.

\begin{figure}[htbp]
\centering
\includegraphics[width=\columnwidth]{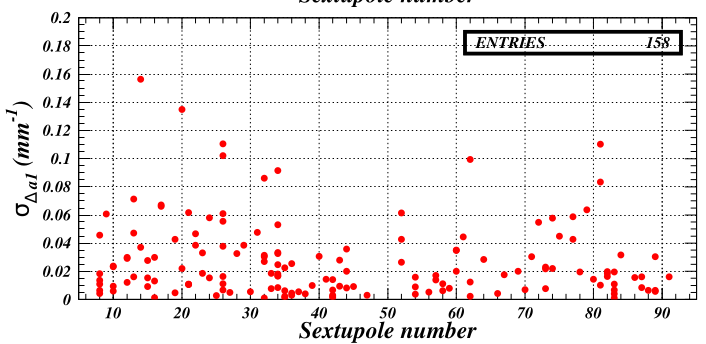}
\caption{Precision values for the skew quadrupole term change $\Delta a_11$ for each sextupole.
  }
   \label{fig:anamain_1_191_5sep2024_11_da1}
\end{figure}

\begin{figure}[htbp]
\centering
\includegraphics[width=\columnwidth]{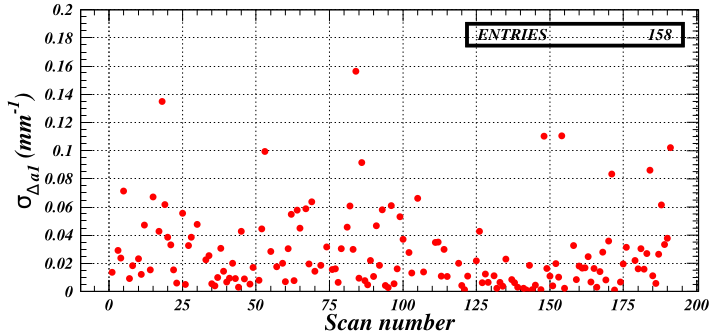}
\caption{Precision values for the skew quadrupole term change $\Delta a_1$ for each $K_2$ scan.
  }
   \label{fig:anamain_1_191_5sep2024_13_da1}
\end{figure}

\begin{figure}[htbp]
\centering
\includegraphics[width=\columnwidth]{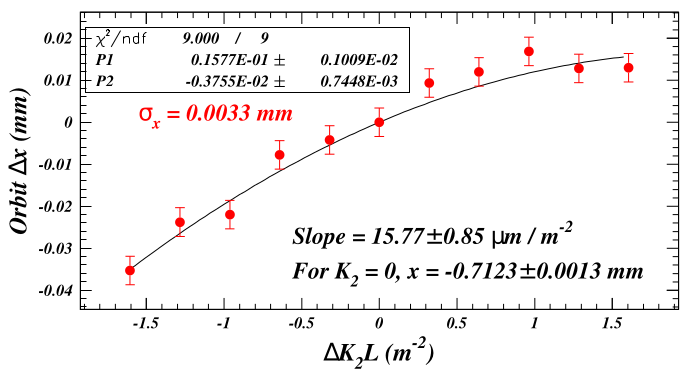}
\caption{The horizontal orbit change $\Delta x$ as a function of sextupole strength change measured in scan 85 for sextupole 10AW. The unconstrained fit value at $K_2 = 0$ has been subtracted. This zero-constrained fit shows the beam position at $K_2 = 0$ to be \mbox{$X_{\rm sext} = -0.7123 \pm 0.0013$~mm.} 
  }
   \label{fig:main_85_40}
\end{figure}

Figures~\ref{fig:anamain_1_191_5sep2024_11_da1}, and~\ref{fig:anamain_1_191_5sep2024_13_da1} 
show the values of the precision in $\Delta a_1$ for each sextupole and each scan.

\subsubsection{Determination of the Beam Position in the Sextupole}

The value for $x$ at $K_2=0$ of \mbox{$-0.7123 \pm 0.0013$~mm} yields a value for the horizontal misalignment \mbox{$X_{\rm offset} = 1.671 \pm 0.0010$~mm.} This means of determining the horizontal misalignment has two advantages over the method presented in Ref.~\cite{Crittenden:IPAC22-MOPOTK040}, which entailed measuring tune changes with sextupole strength at prescribed orbit positions. The first is precision, since the present method uses multi-parameter fits to the entire-ring phase functions and orbit. Secondly, this method can also be used to determine vertical misalignments. Just as Eq.~\eqref{eq:db1} was used above for finding the values of $X_0$, Eq.~\eqref{eq:da1} can be used to find the value for $Y_0$. The corresponding analysis is shown in Fig.~\ref{fig:main_85_23}. The vertical distance of the beam from the center of the sextupole is found to be \mbox{$-0.406 \pm 0.003$~mm.}

\begin{figure}[htbp]
\centering
\includegraphics[width=\columnwidth]{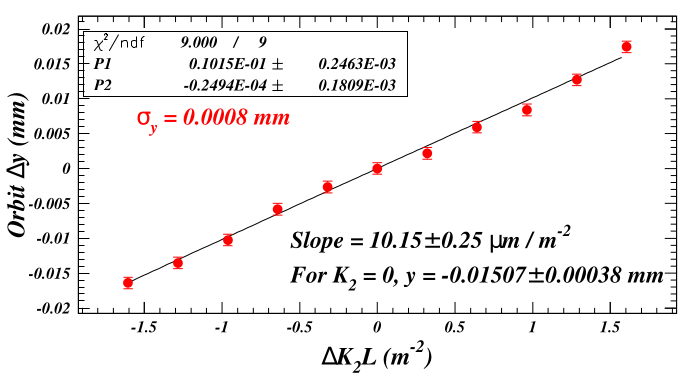}
\caption{The vertical orbit change $\Delta y$ as a function of sextupole strength change $\Delta K_2 L$ measured in scan 85 for sextupole 10AW.
  }
   \label{fig:main_85_42}
\end{figure}

The measurement of the change in the vertical position of the beam in the sextupole is shown in Fig.~\ref{fig:main_85_42}
The value for $y$ at $K_2 = 0$ of \mbox{$0.01507 \pm 0.00038$~mm} yields a value for the vertical misalignment $Y_{\rm offset} = 0.391 \pm 0.003$~mm, the error dominated by the error in the determination of $Y_0$ (see Fig.~\ref{fig:main_85_23}).

\subsection{Results for the Determination of Misalignments}
\label{sec:misalignment_results}
\begin{figure}[htbp]
\centering
\includegraphics[width=\columnwidth]{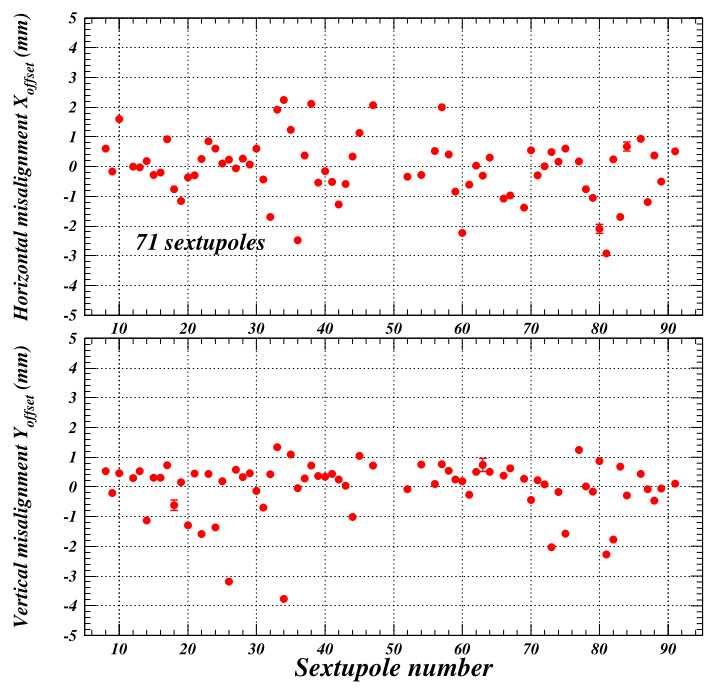}
\caption{ Weighted averages of horizontal and vertical sextupole misalignment values derived from 158 sets of sextupole strength scan data for 71 sextupoles.
  }
   \label{fig:anamain_1_191_5sep2024_40}
\end{figure}

\begin{figure}[htbp]
\centering
\includegraphics[width=\columnwidth]{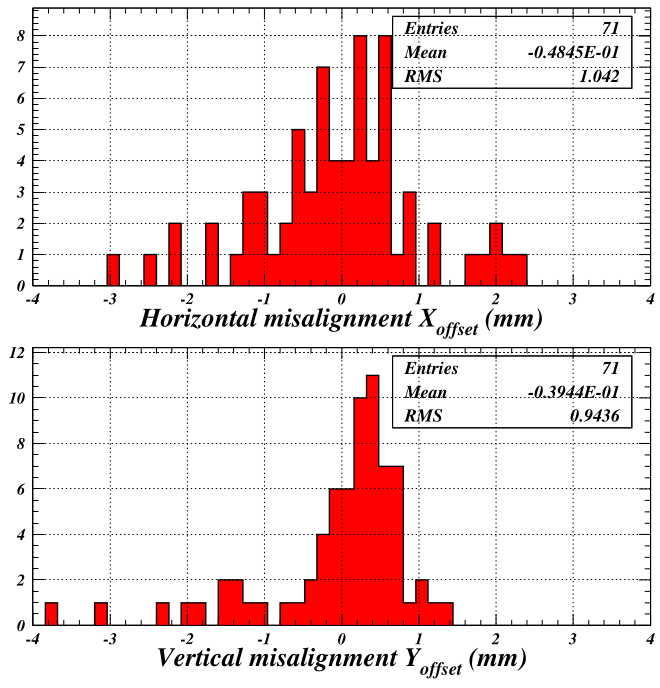}
\caption{Distribution in the weighted averages using multiple measurements of horizontal and vertical sextupole misalignments values.
  }
   \label{fig:anamain_1_191_5sep2024_41}
\end{figure}

\begin{figure}[htbp]
\centering
\includegraphics[width=\columnwidth]{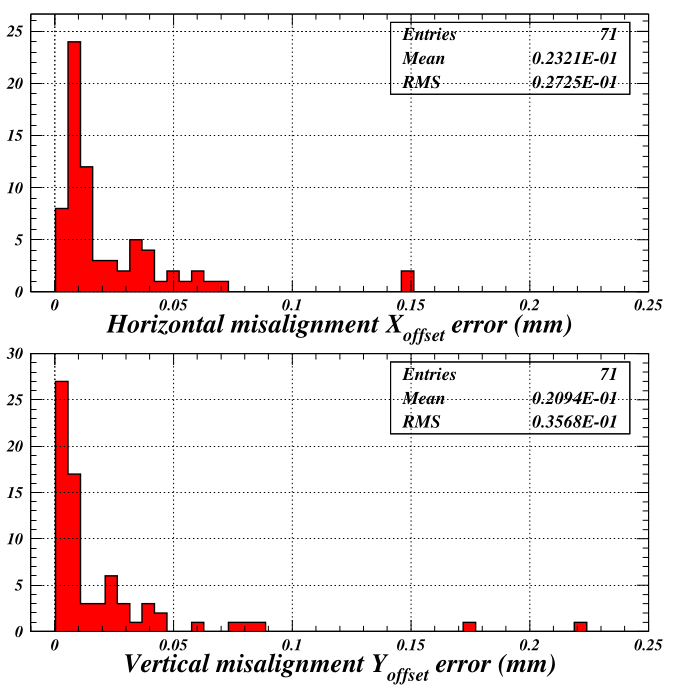}
\caption{Distributions in the uncertainties in the determination of horizontal and vertical sextupole misalignment values.
  }
   \label{fig:anamain_1_191_5sep2024_42}
\end{figure}

We have recorded 158 sextupole strength scans for 71 of the 76 sextupoles in the ring. The error-weighted averages of all measurements are shown in Fig.~\ref{fig:anamain_1_191_5sep2024_40}.
The distributions in the weighted averages of misalignment values and associated uncertainties are shown in
Figs.~\ref{fig:anamain_1_191_5sep2024_41} and~\ref{fig:anamain_1_191_5sep2024_42}.
Typical values for the horizontal misalignments are 1-2~mm. The vertical misalignments
are generally smaller, less than 1~mm, but with a number of exceptions up to 4~mm.
The statistical uncertainties in their determination are typically 0.01 to 0.1~mm. Since beam position dependence on $\Delta K_2$ will lead to deviations from the assumption of linear optics implicit in our derivations, these misalignments must be included in an accurate model of the ring optics.

\subsection{Effects of Sextupole Misalignments}
\label{sec:misalignment_effects}
The procedure to estimate the consequences of the sextupole offsets for the optics was:
\begin{itemize}
  \item
    Perform the optimization on a reference data set as per Sec.~\ref{sec:opt_ref}. The example of one of the three reference measurements for scan 85 was chosen.
  \item
    Set the modeled sextupole offsets to zero sequentially. The optics become unstable if the tunes are not adjusted reset to the measured value after each setting change.
  \item
    Observe the change in the phase, orbit and coupling caused by removing the offsets.
\end{itemize}
The horizontal (vertical) RMS phase deviation from the design value around the ring was found to increase to 1.74~(1.45)~degrees. The coupling RMS increased to 0.044. These errors are easily compensated by the quadrupole and skew quadrupole magnets within their operating ranges. The RMS changes in the horizontal and vertical orbits was less than 0.1~mm. The horizontal (vertical) chromaticity changed from 0.980 to 1.09 (1.00 to 0.95).

It has also been shown via modeling that the effect of the sextupole offsets on dynamic aperture is minor~\cite{crittenden_15feb2023}.

\section{Beam Size Calculations}
\label{sec:beamsize}
\subsection{Derivations}
Here we generalize the derivation presented in Sec.~\ref{sec:introduction} to the case of non-zero initial sextupole setting, initially in a single transverse dimension. The variation of a sextupole strength by an amount $\Delta K_2L $ in a storage ring introduces
\begin{enumerate}
\item
a quadrupole kick $\Delta K_1L$
\begin{equation}
  \label{eq:quadkick}
  \Delta K_1L = X_0 \; \Delta K_2L + \left( K_2L + \Delta K_2L \right) \Delta x, 
\end{equation}
and
\item
a horizontal angle change $\Delta p_X$
\begin{eqnarray}
  \label{eq:dpx}
  \Delta p_X &=& \frac{1}{2} \left( X^2_0 + \sigma^2_{\rm X} \right)\Delta K_2L \notag \\
    &+& \frac{1}{2} \left(2 \; X_0 \; \Delta x +  \Delta x^2 \right)  \left( K_2L + \Delta K_2L \right),
\end{eqnarray}
\end{enumerate}
where $L$ is the length of the sextupole, $\Delta x$ is the change of the
beam position from its original position relative to the center of the sextupole $X_0$, and $\sigma_x$ is the beam size.

The two equations permit the elimination of the unknown value $X_0$ to determine the beam size from the measured
values of $\frac{\Delta K_1L}{\Delta K_2L}$ and $\frac{\Delta x}{\Delta K_2L}$:
\begin{eqnarray}
  \sigma^2_{\rm X} &=& \frac{4\tan (\pi Q)}{\beta}  \frac{\Delta x}{\Delta K_2L} -  \left( \frac{\Delta K_1L}{\Delta K_2L} \right)^2 \notag \\
  &+& (K_2L \frac{\Delta x}{\Delta K_2L})^2  \left( 1+\frac{\Delta K_2L}{K_2L} \right),
\end{eqnarray}
where the  values prior to the variation are the tune~$Q$, the sextupole strength~$K_2L$, and the twiss function $\beta$ at the sextupole.
We note that no terms have been neglected in this derivation.

The judicious choice for the initial value of the sextupole strength $K_2L=0$, together with
the fact that a kick $\Delta p_X$ causes a closed orbit change at the location of the kick $\Delta x$ given by
\begin{equation}
  \Delta x = \frac{\beta \;\cot (\pi Q)}{2}  \Delta p_X,
\end{equation}
results in the simple relationship
\begin{equation}
  \sigma^2_{\rm X} = 2 \frac{\Delta p_X}{\Delta K_2L} -  \left( \frac{\Delta K_1L}{\Delta K_2L} \right)^2.
\end{equation}

Extending the analysis to the two transverse dimensions~(\cite{Crittenden:IPAC21-MOPAB254}, ~\cite{Crittenden:IPAC22-MOPOTK040}), we
derive the quadrupole kick $\Delta K_1L$ and the angle changes $\Delta p_X$ and $\Delta p_Y$ from a change in sextupole strength $\Delta K_2L$ using the sextupole field components $\frac{qL}{p_0}B_{\rm x} =  K_2L xy$ and $\frac{qL}{p_0}B_{\rm y} = \frac{1}{2} K_2L (x^2 - y^2 )$, we obtain three equations with four unknowns:
   \begin{equation}
     \Delta   K_1L = \Delta  K_2L \left( X_0 + \Delta x \right)
   \end{equation}
   \begin{equation}
     \Delta p_Y = \Delta  K_2L \left( X_0 + \Delta  x \right) \left( Y_0 + \Delta  y \right)
   \end{equation}
   \begin{equation}
     2\; \Delta p_X = \Delta  K_2L \left[  \left( \frac{\Delta p_Y}{\Delta  K_2L} \right)^2 \left( \frac{\Delta   K_1L}{\Delta  K_2L} \right)^{-2} + \sigma^2_{\rm Y}
       - \left( \frac{\Delta   K_1L}{\Delta  K_2L} \right)^2  - \sigma^2_{\rm X}  \right]
   \end{equation}
   We note that these quantities are differences, not differentials. The equations are exact; there is no expansion. Typical vertical beam sizes are approximately 0.05~mm, too small to measure using our method. We neglect their small contribution the calculation of horizontal beam size, which is typically 1-2-mm.

Assuming initial $K_2L = 0$ and including all terms:\\
   \begin{equation}
  \sigma^2_{\rm X} - \sigma^2_{\rm Y} = - 2\; \frac{\Delta p_X}{\Delta K_2L} + \left( \frac{\Delta p_Y}{\Delta  K_2L} \right)^2 \left( \frac{\Delta K_1L}{\Delta K_2L} \right)^{-2} - \left( \frac{\Delta K_1L}{\Delta K_2L}  \right)^2
  \end{equation}

Including only terms linear in $\Delta K_2L$, we have:\\
\begin{equation}
  \label{eq:size}
  \sigma^2_{\rm X} - \sigma^2_{\rm Y} = - 2\; \frac{\Delta p_X}{\Delta  K_2L} + Y_0^2 - X_0^2,
  \end{equation}
where $X_0$, $Y_0$ is the initial position of the beam relative to the center of the sextupole.  The is the two-dimensional generalization of Eq.~5 in our IPAC21 paper~\cite{Crittenden:IPAC21-MOPAB254}.

\subsection{Modeling for Beam Size Measurement}
\begin{figure}[htbp]
\centering
\includegraphics[width=\columnwidth]{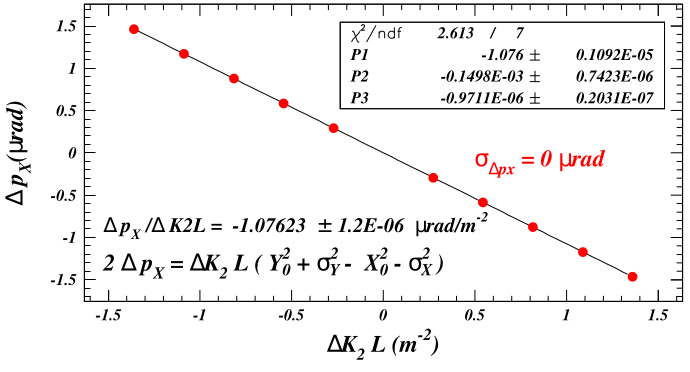}
\caption{The angle change $\Delta p_X$ as a function of the sextupole 10AW strength change $\Delta K_2$, calculated by tracking 2000 beam particles through the CESR design lattice. The value for the linear coefficient -1.076~$\mu$rad/m$^{-2}$ gives a reconstructed beam size value of 1.070~mm via Eq.~\ref{eq:sigma}. The value of the horizontal RMS spread of the tracked beam particles at the sextupole is 1.062~mm. The uncertainties given the for the coefficients are due to machine accuracy in the modeling. The weighting to obtain $\chi^2$/NDF=1 also results in rounding errors at machine accuracy.
  }
   \label{fig:ana_k2scan_10_10aw_16}
\end{figure}

For the beam size calculation, we need the angle change in the sextupole $\Delta p_X$ as well as the quadrupole term $X_0$ calculated in Sec.~\ref{sec:model_alignment}. This is shown in Fig.~\ref{fig:ana_k2scan_10_10aw_16}.
The value for the linear coefficient -1.076~$\mu$rad/m$^{-2}$ gives a reconstructed beam size value of 1.070~mm via Eq.~\ref{eq:sigma}. The value of the horizontal RMS spread of the tracked beam particles at the sextupole is consistent at 1.062~mm.

\begin{figure}[htbp]
\centering
\includegraphics[width=\columnwidth]{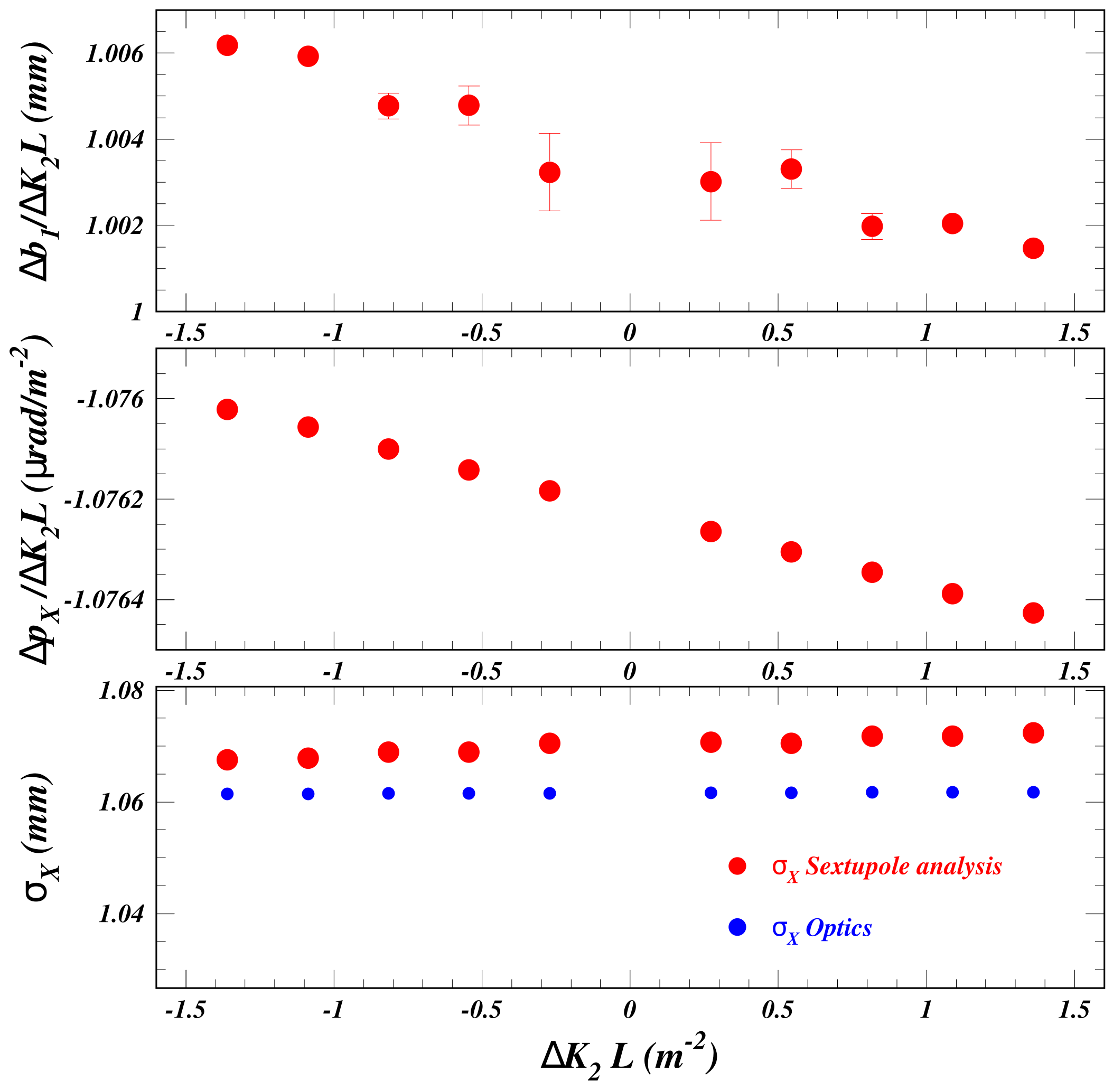}
\caption{The values of $\Delta b_1 / \Delta K_2 L$ and $\Delta p_X / \Delta K_2 L$ for the ten values of $\Delta K_2 L$ are shown in the top two plots. The third plot shows in red the dependence of the reconstructed beam size on the $K_2$ change. The blue points show the horizontal RMS spread of the tracked beam particles at the sextupole. The value reconstructed for the beam size shows little dependence on the magnitude of the change in $K_2$.
  }
   \label{fig:ana_k2scan_10_10aw_19}
\end{figure}

The quadrupole term was also calculated from the difference phase function, as described in Sec.~\ref{sec:quad_fit} for the analysis of the $K_2$ scan data. The resulting values of $\Delta b_1 / \Delta K_2 L$ and $\Delta p_X / \Delta K_2 L$ for the ten values of $\Delta K_2 L$ are shown in Fig.\ref{fig:ana_k2scan_10_10aw_19}
Nonlinear effects are observed to be small relative to the accuracy of the beam size calculation. The reconstructed beam size shows little dependence on the magnitude of the change in $K_2$.

\begin{figure}[htbp]
\centering
\includegraphics[width=\columnwidth]{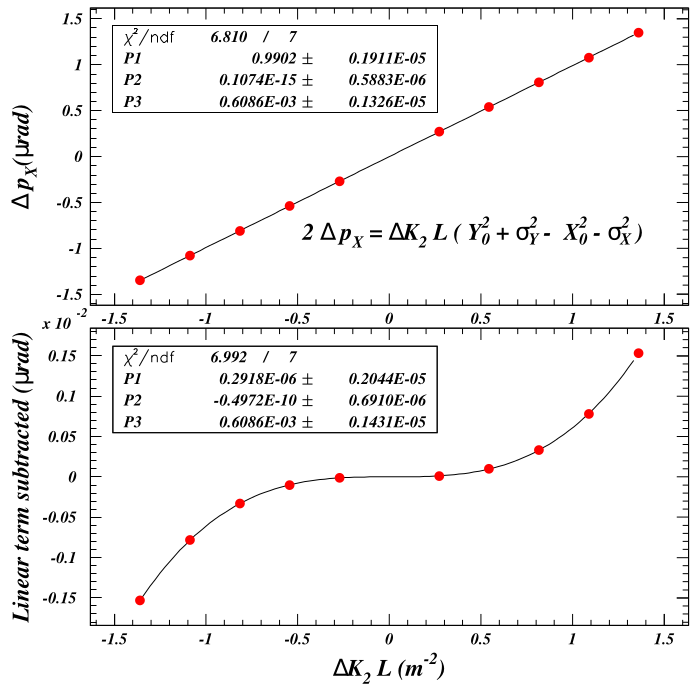}
\caption{Modeled results for the angle change $\Delta p_X$ dependence on sextupole strength change for the case with no sextupole offset. The linear coefficient value of -0.9902~$\mu$rad/m$^{-2}$ gives the expected value for the beam size of 1.4~mm via $\sigma^2_x = -2 \Delta p_X / \Delta K_2 L$. The cubic coefficient arises from the term $\frac{1}{2} \left( \Delta x \right)^2 \Delta K_2 L$ as seen in Eq.~\ref{eq:dpx}. The uncertainties given the for the coefficients are due to machine accuracy in the modeling.
  }
   \label{fig:ana_k2scan_1_5_0_b0_19}
\end{figure}

\begin{figure}[htbp]
\centering
\includegraphics[width=\columnwidth]{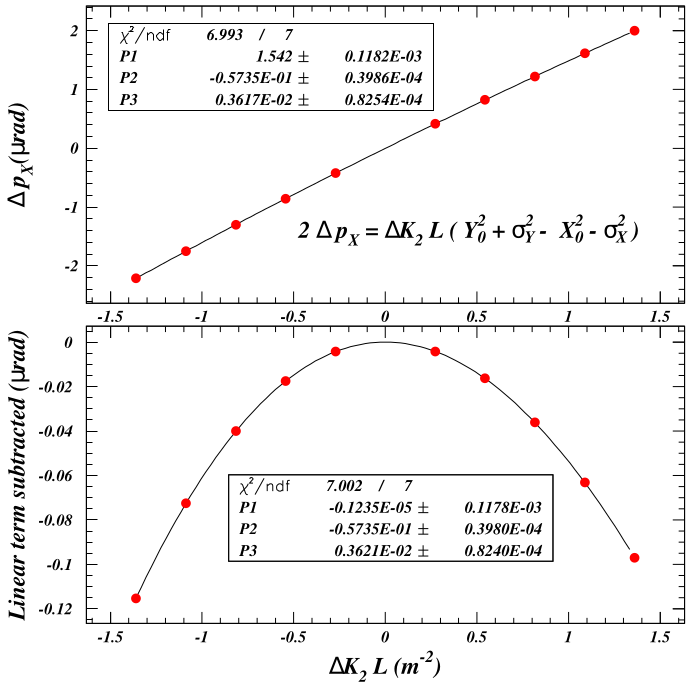}
\caption{Modeled results for the angle change $\Delta p_X$ dependence on sextupole strength change for the case with a sextupole offset of 1~mm.
The linear coefficient value of -1.542~$\mu$rad/m$^{-2}$ gives the expected value for the beam size of 1.4~mm via $\sigma^2_x = -2 \Delta p_X / \Delta K_2 L - X_0^2$. For this value of $X_0 = 1$~mm, the quadratic term in Eq.~\ref{eq:dpx}, $X_0 \; \Delta x \; \Delta K_2 L$, dominates over the cubic term. The uncertainties given the for the coefficients are due to machine accuracy in the modeling.
  }
   \label{fig:ana_k2scan_1_5_001_b0_19}
\end{figure}

Following the above study using beam particle tracking, we realized that a simpler test of the model can be made by superposing on the sextupole the dipole term expected from the beam size, $b_0 = \frac{1}{2} \Delta K_2 \; L \; \sigma^2_{\rm X}$, and reconstructing the beam size by our method using the tune change and the angle change caused by the change in the closed orbit. This method was tested using a toy FODO lattice with a single sextupole. Figures~\ref{fig:ana_k2scan_1_5_0_b0_19} and~\ref{fig:ana_k2scan_1_5_001_b0_19}
show the modeled results for the $\Delta p_X$ dependence on sextupole strength change for the cases with no sextupole offset and with a sextupole offset of 1~mm.

\begin{figure}[htbp]
\centering
\includegraphics[width=\columnwidth]{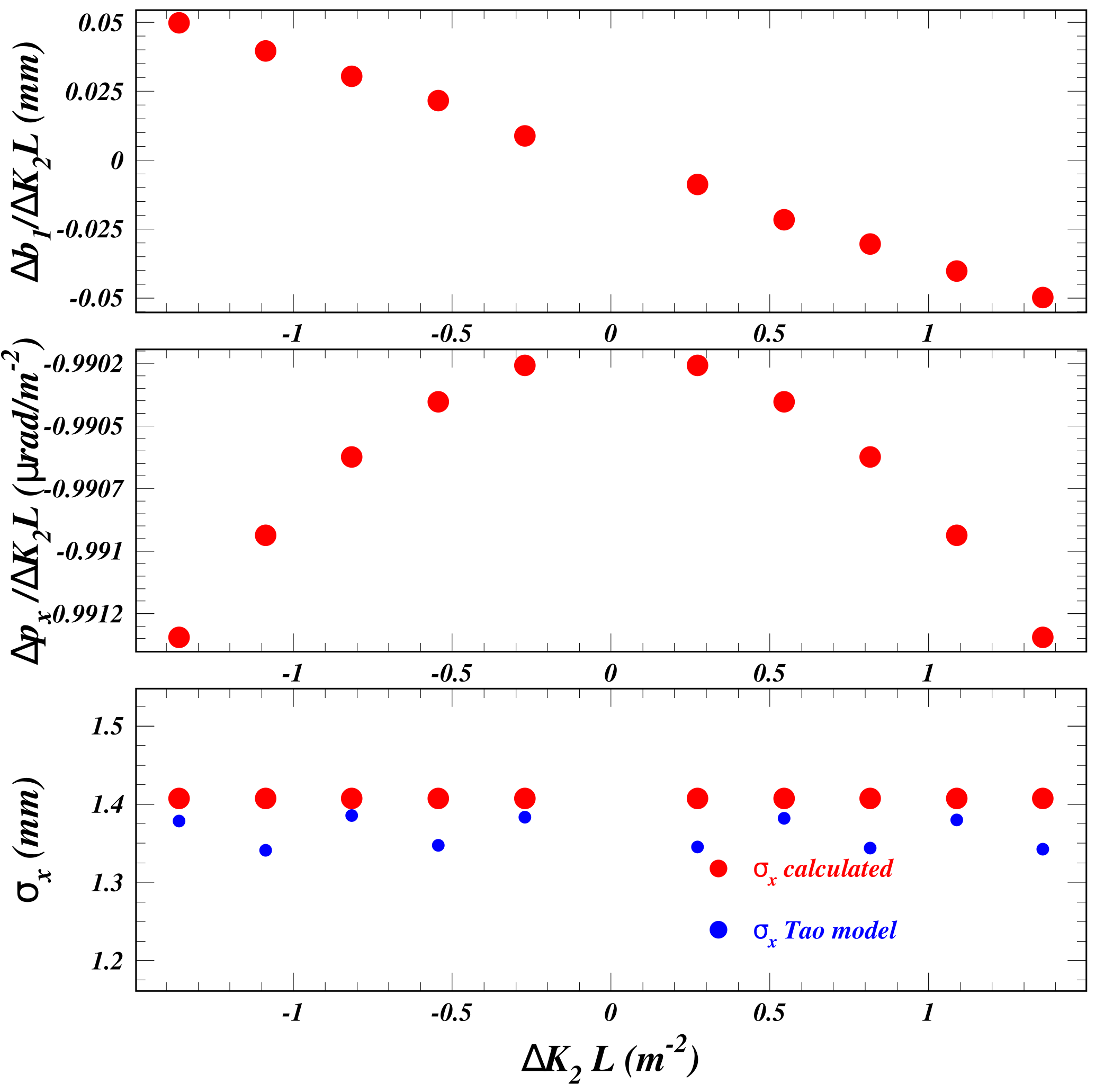}
\caption{Modeled results for the beam size calculation dependence on sextupole strength change for the case of no sextupole offset. The dependence for both the quadrupole slope and the horizontal angle slope are very weak.
  }
   \label{fig:ana_k2scan_1_5_0_b0_17}
\end{figure}

\begin{figure}[htbp]
\centering
\includegraphics[width=\columnwidth]{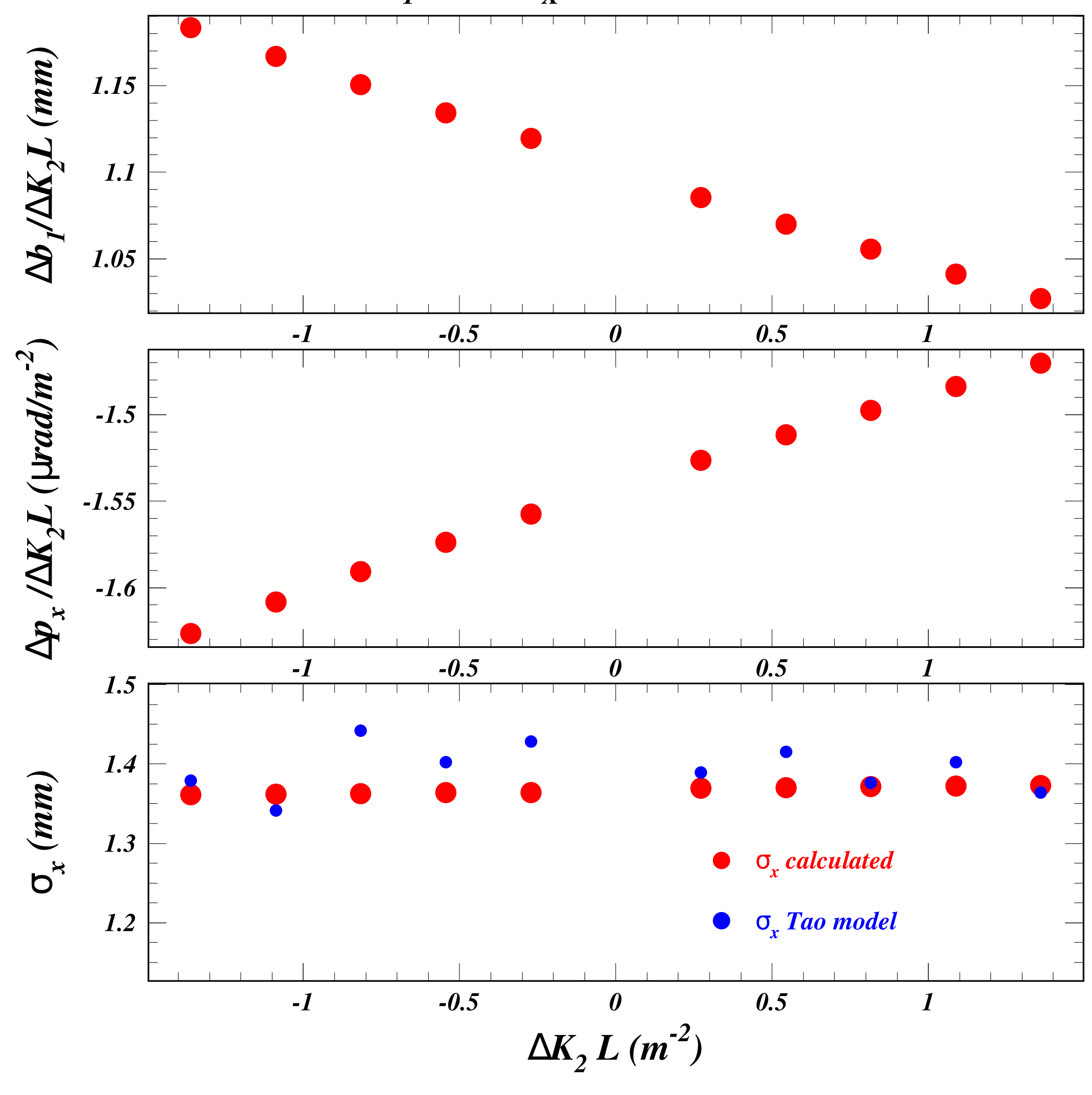}
\caption{Modeled results for the beam size calculation dependence on sextupole strength change for the case with a sextupole offset of 1~mm.
  }
   \label{fig:ana_k2scan_1_5_001_b0_17}
\end{figure}

The modeled results for the beam size calculation are shown in Figs.~\ref{fig:ana_k2scan_1_5_0_b0_17} and~\ref{fig:ana_k2scan_1_5_001_b0_17}.

\subsection{Analysis Procedure for the Difference Angle Change Functions}
\label{sec:analysis_orbit}
\subsubsection{Scatter Plots}
\begin{figure}[htbp]
\centering
\includegraphics[width=\columnwidth]{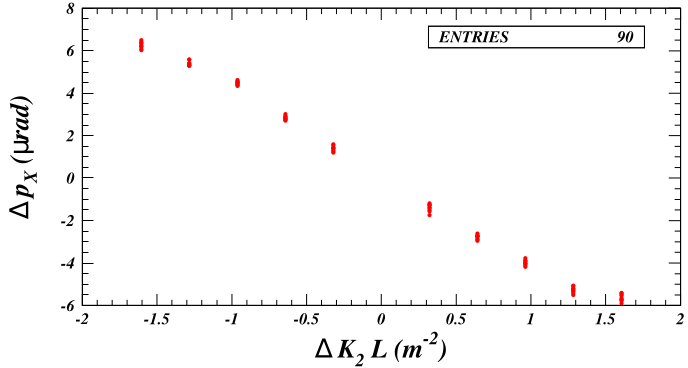}
\caption{Scatter plots for the three difference measurements at eleven $K_2$ values for the horizontal angle change $\Delta p_{\rm x}$ induced by the change in sextupole strength $\Delta K_2$. This example is scan 85 for sextupole 10AW.
  }
   \label{fig:main_85_13_dpx}
\end{figure}

Figure~\ref{fig:main_85_13_dpx}
shows the scatter plot in the horizontal orbit angle change in the sextupole 10AW for the three measurements at each of the eleven $K_2$ settings in scan~85. The repeatability is observed to be less than a few tenths of a degree for a full range of angle change of $\pm 6$~degrees providing for a precise measurement of the linear term in the polynomial fit described in Sec.~\ref{sec:angle_polyfit}. The linear coefficient is the most consequential parameter in the calculation of beam size.

\begin{figure}[htbp]
\centering
\includegraphics[width=\columnwidth]{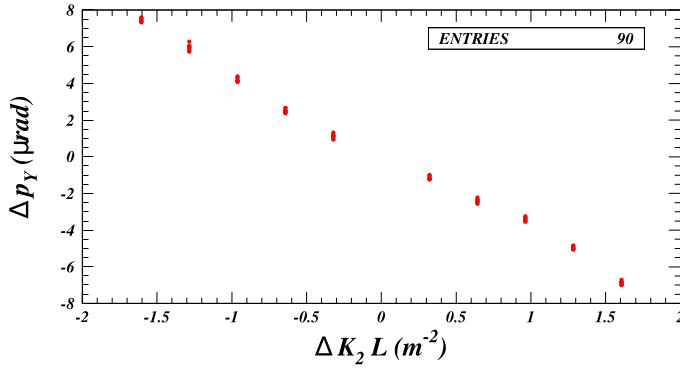}
\caption{Scatter plots for the three difference measurements at eleven $K_2$ values for the vertical angle change $\Delta p_{\rm y}$ induced by the change in sextupole strength $\Delta K_2$. This example is scan 85 for sextupole 10AW.
  }
   \label{fig:main_85_14_dpy}
\end{figure}

The corresponding plots for the vertical angle change are shown in Fig.~\ref{fig:main_85_14_dpy},
exhibiting a data quality similar to that observed for the horizontal angle change measurement.

\subsubsection{Polynomial Fits and Error Analysis}
\label{sec:angle_polyfit}
\subsubsection{Horizontal Angle Change}
\begin{figure}[htbp]
\centering
\includegraphics[width=\columnwidth]{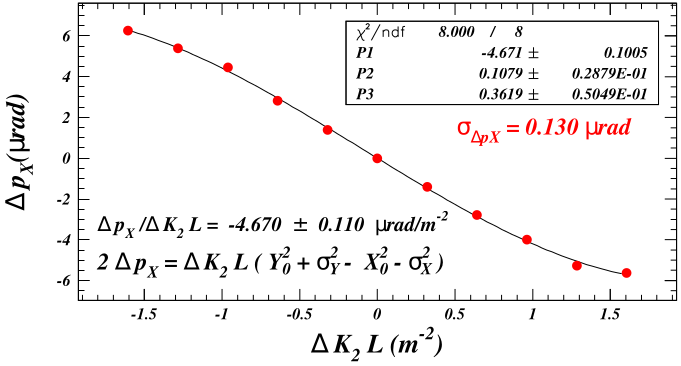}
\caption{Polynomial fit results for the horizontal angle change $\Delta p_{\rm x}$ induced by the change in sextupole strength $\Delta K_2$.
  }
   \label{fig:main_85_21}
\end{figure}

Figure~\ref{fig:main_85_21}
 show the results for horizontal angle change dependence on the $K_2$ change for the example of the $K_2$ scan for sextupole 10AW. The data show a clear need for a cubic term, as expected from Eqs.~\ref{eq:dpy} and~\ref{eq:intro_dpx}.
The lack of necessity for a quartic term at our level of precision encourages the approximation that nonlinear effects are small.

Given the values of $X_0$ and $Y_0$ obtained for this scan in Secs.~\ref{sec:quad_fit} and~\ref{sec:skew_quad_fit} of \mbox{$-2.383 \pm 0.010$~mm} and \mbox{$0.406 \pm 0.003$~mm}, this value
\begin{equation*}
  \Delta p_X / \Delta K_2 L = -4.670 \pm 0.110~\mu{\rm rad}/{\rm m}^{-2}
\end{equation*}
results in a beam size determination of $1.956 \pm 0.053$~mm, much larger than the value expected from the optics, 1.09~mm. We will see below that this is a problem common among the $K_2$ scans. The precision in the determination of this linear coefficient is good enough to measure a 1-mm beam size with 10\% accuracy, but the systematic error is much larger.

\begin{figure}[htbp]
\centering
\includegraphics[width=\columnwidth]{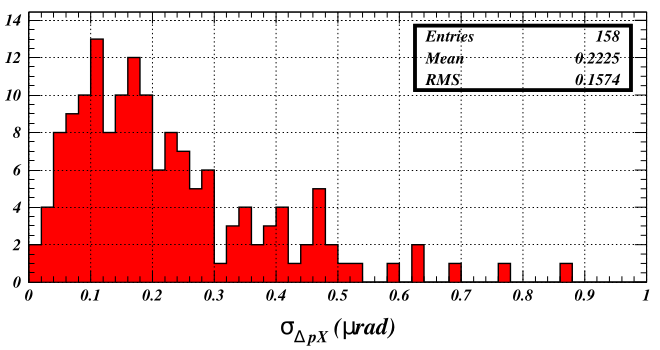}
\caption{Distribution in precision values for the horizontal orbit angle change $\Delta p_{\rm X}$ for all scans.
  }
   \label{fig:anamain_1_191_5sep2024_10_dpx}
\end{figure}

\begin{figure}[htbp]
\centering
\includegraphics[width=\columnwidth]{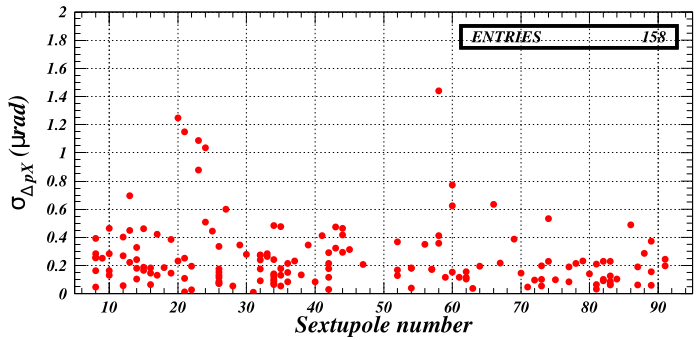}
\caption{Precision values for the horizontal orbit angle change $\sigma_{\Delta p_{\rm X}}$ for each sextupole.
  }
   \label{fig:anamain_1_191_5sep2024_12_dpx}
\end{figure}

\begin{figure}[htbp]
\centering
\includegraphics[width=\columnwidth]{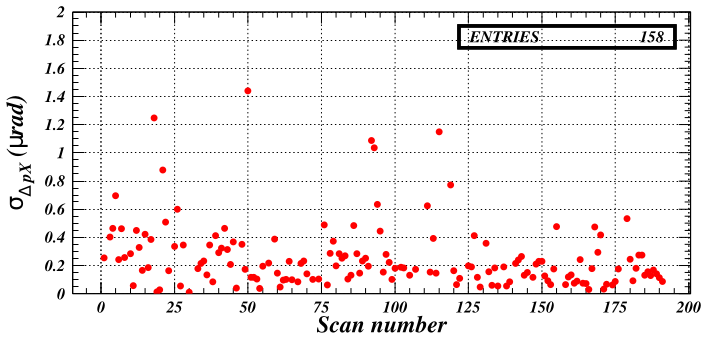}
\caption{Precision values for the horizontal orbit angle change $\sigma_{\Delta p_{\rm X}}$ for each $K_2$ scan.
  }
   \label{fig:anamain_1_191_5sep2024_14_dpx}
\end{figure}

The distribution in precision values for the horizontal change $\Delta p_{\rm X}$ for all scans is shown in Fig.~\ref{fig:anamain_1_191_5sep2024_10_dpx}.
Typical values are 0.1-0.2~$\mu$rad, which is sufficiently accurate for a precision of about 30\% for a 1-mm beam size. Figures~\ref{fig:anamain_1_191_5sep2024_12_dpx} and ~\ref{fig:anamain_1_191_5sep2024_14_dpx}.
show the precision values for each sextupole and each scan.

\subsubsection{Vertical Angle Change}
\begin{figure}[htbp]
\centering
\includegraphics[width=\columnwidth]{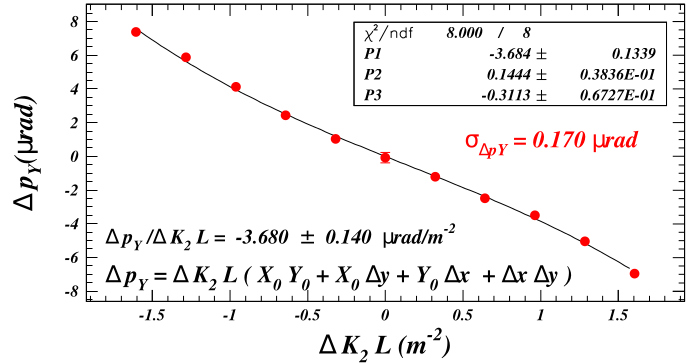}
\caption{Polynomial fit results for the vertical orbit angle change $\Delta p_{\rm Y}$ induced by the change in sextupole strength $\Delta K_2$ for the case of scan~85.
  }
   \label{fig:main_85_25}
\end{figure}

\begin{figure}[htbp]
\centering
\includegraphics[width=\columnwidth]{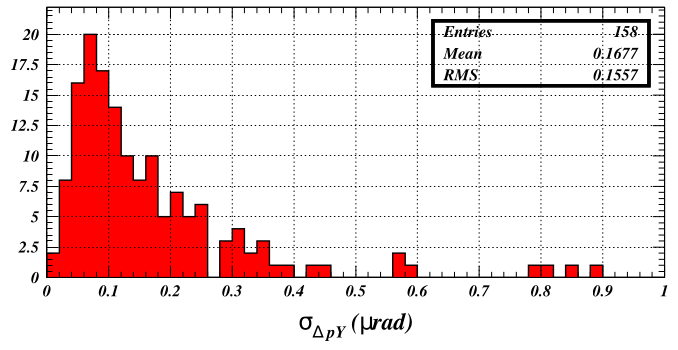}
\caption{Distribution in precision values for the vertical angle change $\sigma_{\Delta p_{\rm Y}}$ for all scans.
  }
   \label{fig:anamain_1_191_5sep2024_10_dpy}
\end{figure}

Figure~\ref{fig:main_85_25} 
shows that the precision in determining the vertical angle change is similar to that for the horizontal angle change. The need for a cubic term (see Eq.~\ref{eq:dpy}) is clearly determined at $-0.311 \pm 0.067 \mu$rad. The precision in determining $\Delta p_Y$ is found to be 0.17~$\mu$rad. This is a fairly typical value for all scans, as is seen in Fig.~\ref{fig:anamain_1_191_5sep2024_10_dpy}.
The average value is 0.167~$\mu$rad and the RMS value of the distribution is 0.154~$\mu$rad.

\begin{figure}[htbp]
\centering
\includegraphics[width=\columnwidth]{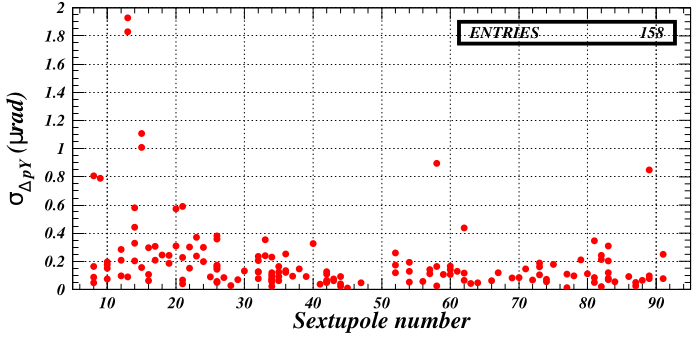}
\caption{Precision values for the vertical orbit angle change  $\sigma_{\Delta p_{\rm Y}}$ for each sextupole.
  }
   \label{fig:anamain_1_191_5sep2024_12_dpy}
\end{figure}

\begin{figure}[htbp]
\centering
\includegraphics[width=\columnwidth]{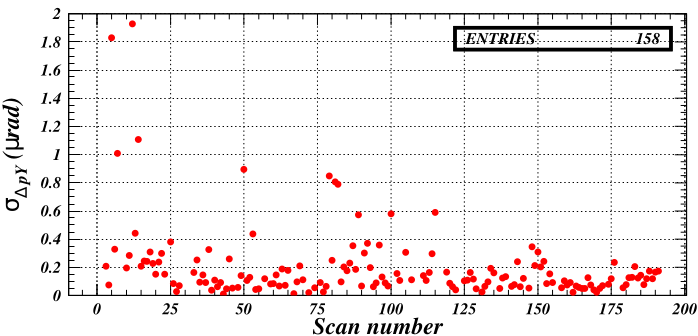}
\caption{Precision values for the vertical orbit angle change $\sigma_{\Delta p_{\rm Y}}$ for each $K_2$ scan.
  }
   \label{fig:anamain_1_191_5sep2024_14_dpy}
\end{figure}

The values in the precision of the $\Delta p_Y$ determination for each sextupole and each scan are shown in Figs.~\ref{fig:anamain_1_191_5sep2024_12_dpy} and~\ref{fig:anamain_1_191_5sep2024_14_dpy}.

\subsubsection{Error Analysis for Beam Size Calculation}
We recall Eq.~\ref{eq:sigma}, neglecting the small contribution from the vertical beam size,
\begin{equation}
  \label{eq:sigx}
  \sigma^2_{\rm X} = - 2\; \frac{\Delta  p_{\rm X}}{\Delta K_2L} + Y_0^2 - X_0^2. 
   \end{equation}
   So the uncertainty in the squared beam size calculation, $\delta_{\sigma^2_X}$, is then given by
   \begin{equation}
     \delta_{\sigma^2_{\rm X}} =  4 \left[ \left( \delta_{\frac{\Delta  p_{\rm X}}{\Delta K_2L}} \right)^2 + \left( Y_0 \; \delta_{Y_0} \right)^2 + \left( X_0 \; \delta_{X_0} \right)^2  \right],
   \end{equation}
where $\delta_{\frac{\Delta  p_{\rm X}}{\Delta K_2L}}$, $\delta_{Y_0}$, and $\delta_{X_0}$ are the uncertainties in the the measured quantities.
The uncertainty in the beam size is given by
   \begin{equation}
     \delta_{\sigma_{\rm X}} = \frac{\delta_{\sigma^2_{\rm X}}}{2\;\sigma_{\rm X}}
   \end{equation}

\begin{figure}[htbp]
\centering
\includegraphics[width=\columnwidth]{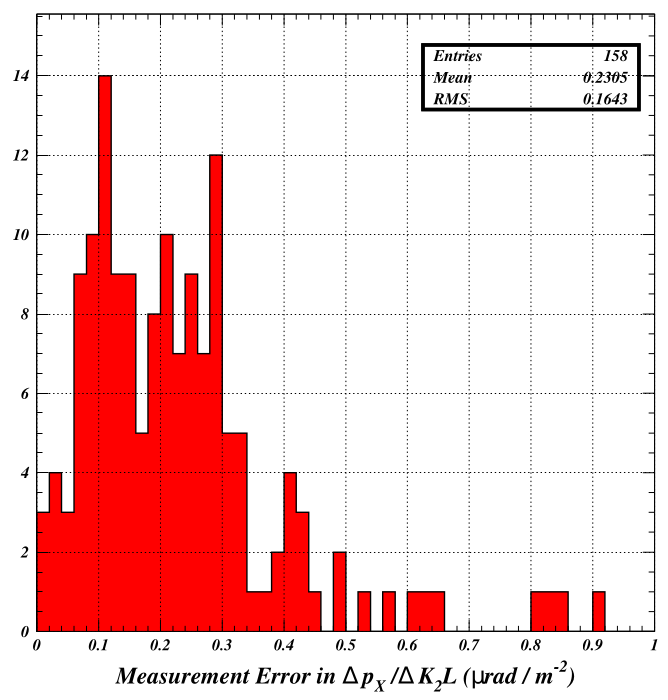}
\caption{Distribution in the precision values for the horizontal angle change slope $\frac{\Delta  p_{\rm X}}{\Delta K_2L}$.
  }
   \label{fig:anamain_1_191_5sep2024_1}
\end{figure}

\begin{figure}[htbp]
\centering
\includegraphics[width=\columnwidth]{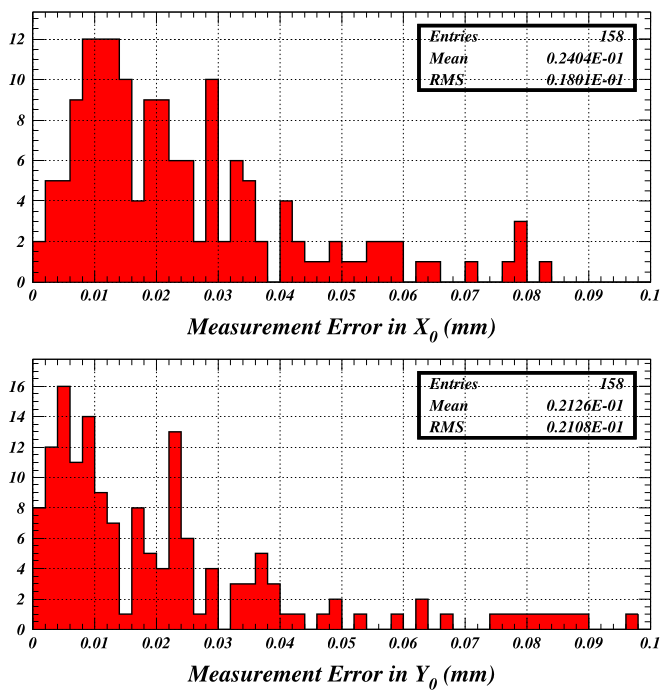}
\caption{Distribution in precision values for $X_0$ and $Y_0$.
  }
   \label{fig:anamain_1_191_5sep2024_2}
\end{figure}

   See Figs.~\ref{fig:anamain_1_191_5sep2024_1} and~\ref{fig:anamain_1_191_5sep2024_2} for the three contributions to the precision of the beam size calculation.

Since the uncertainty in the linear term in horizontal angle change dependence on the sextupole strength change is typically less than 0.3~$\mu{\rm rad}/{\rm m}^{-2}$, the resulting contribution to the error in a 1~mm beam size measurement is 30\%. To this precision the analysis described above can be considered acceptable for measuring the beam size at each sextupole magnet in CESR. The contributions by the uncertainties in $X_0$ and $Y_0$ of typically less than 0.05~mm are small in comparison.

\begin{figure}[htbp]
\centering
\includegraphics[width=\columnwidth]{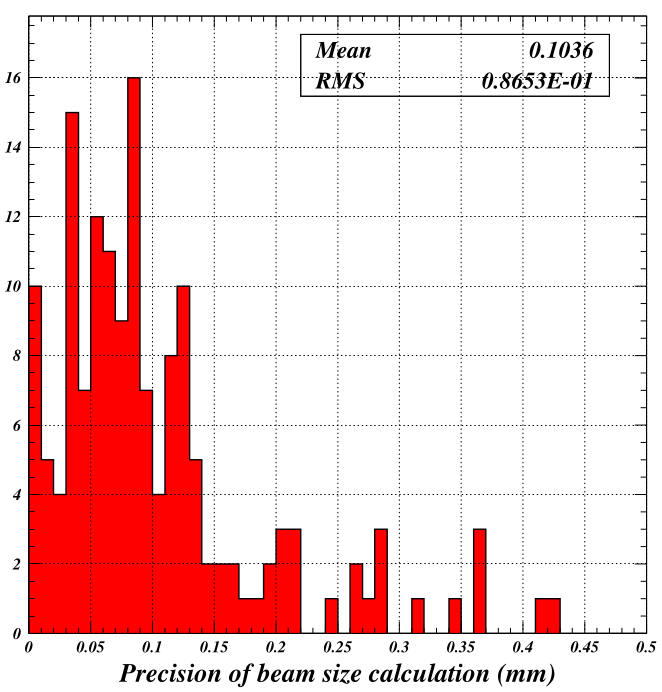}
\caption{Distribution in beam size measurement precision values propagated from the three contributions to its calculation.
  }
   \label{fig:anamain_1_191_5sep2024_47}
\end{figure}

Figure~\ref{fig:anamain_1_191_5sep2024_47}
shows the distribution of beam size precision values. Typical values are \mbox{0.05 to 0.15~mm.}

\subsubsection{Results of the Beam Size Calculation}
\begin{figure}[htbp]
\centering
\includegraphics[width=\columnwidth]{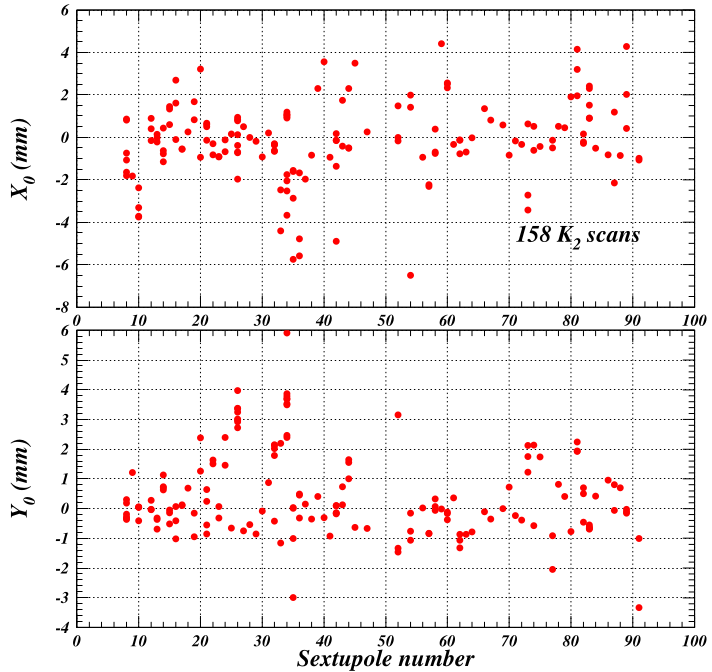}
\caption{Results of the fits to difference functions for the linear terms in the dependence of quadrupole kick $b_1$ and skew quadrupole kick $a_1$ on sextupole strength. These terms are equal to the horizontal (vertical) distance of the beam from the center of the sextupole $X_0$ ($Y_0$). These values, together with the linear term in the horizontal angle change dependence on sextupole strength, suffice to calculate the horizontal beam size for each scan. 
  }
   \label{fig:anamain_1_191_5sep2024_22}
\end{figure}

\begin{figure}[htbp]
\centering
\includegraphics[width=\columnwidth]{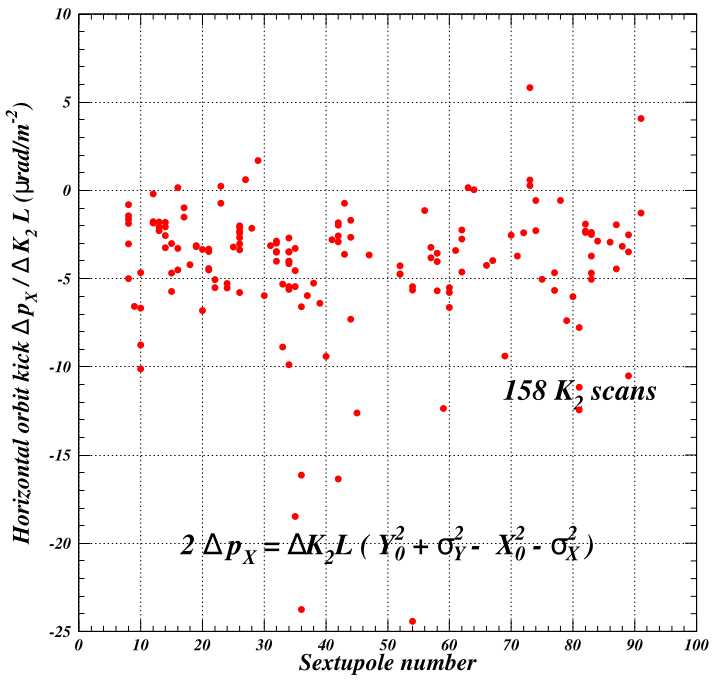}
\caption{Results of the fits to difference functions for the linear term in the horizontal angle change dependence on sextupole strength for all scans. These can differ for each scan, since the beam position relative to the center of the sextupole differs. However, in general, these terms are found to be too negative to be consistent with the beam size.
  }
   \label{fig:anamain_1_191_5sep2024_18}
\end{figure}

Recalling the equation for the beam size (Eq.~\ref{eq:sigx}), we plot the determinations of the three contributions $X_0$, $Y_0$ and $\Delta p_X / \Delta K_2 L$ in Figs.~\ref{fig:anamain_1_191_5sep2024_22} and \ref{fig:anamain_1_191_5sep2024_18}
for each sextupole. Multiple scans were made for a number of sextupoles. The calculated values may be different for each scan, since the beam position relative to the sextupole center may differ from scan to scan. For the same beam size and $Y_0$ value, a larger value of $X_0$ requires a larger horizontal angle change with $\Delta K_2$.

\begin{figure}[htbp]
\centering
\includegraphics[width=\columnwidth]{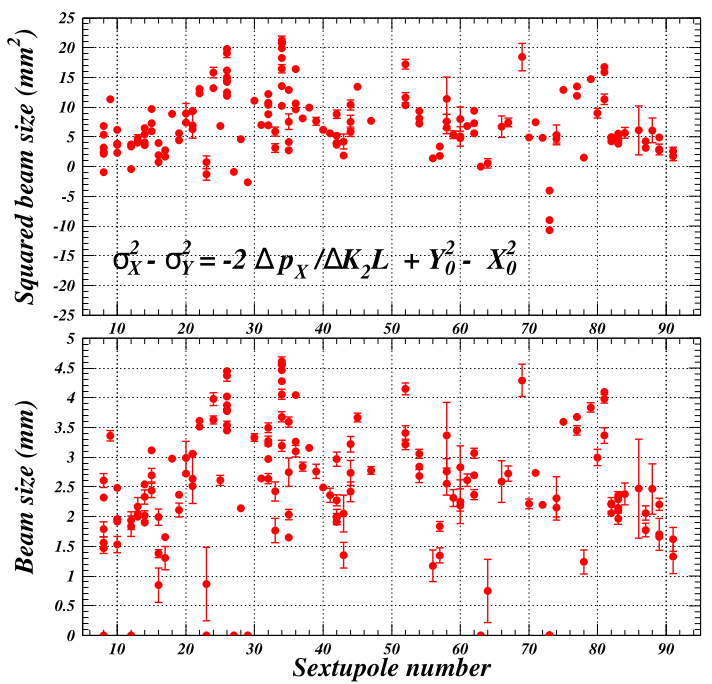}
\caption{Results for the beam size calculation for each sextupole. Note the multiple entries and repeatability. The beam size is shown as zero for the cases where the squared beam size is negative.
  }
   \label{fig:anamain_1_191_5sep2024_45}
\end{figure}

\begin{figure}[htbp]
\centering
\includegraphics[width=\columnwidth]{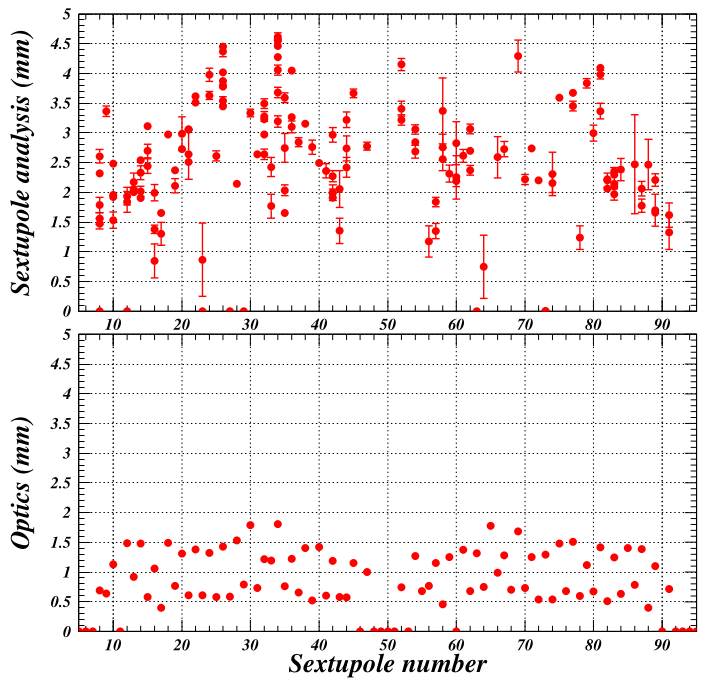}
\caption{Comparison of the beam size calculations to the beam size values expected from the optical functions and dispersion.
  }
   \label{fig:anamain_1_191_5sep2024_48}
\end{figure}

Figure~\ref{fig:anamain_1_191_5sep2024_45}
shows the values for the squared beam size and the beam size for all scans for each sextupole magnet. Note that the finite measurement precision can result in negative values for the squared beam size. There are then shown as zeroes in the plot of beam size. In general, we see that the values for $\Delta p_X / \Delta K_2 L$ tend to be too negative to be consistent with the value for the beam size expected from the optics, as shown in Fig.~\ref{fig:anamain_1_191_5sep2024_48}.

\begin{figure}[htbp]
\centering
\includegraphics[width=\columnwidth]{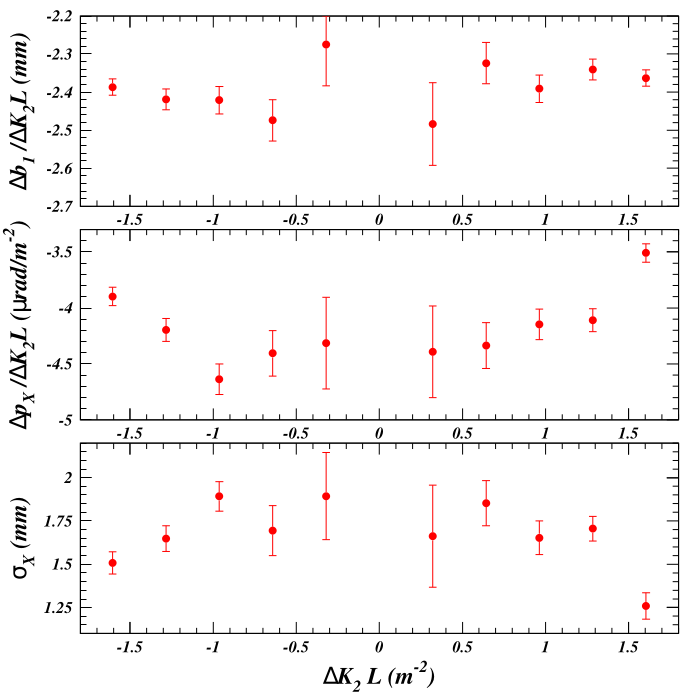}
\caption{The dependence of the contributions to the beam size calculation on the sextupole strength change $\Delta K_2$. No dependence for $\Delta b_1 / \Delta K_2L$ is observed within uncertainties. There is a clear dependence on $\Delta K_2$ observed for $\Delta p_X / \Delta K_2L$. It results in the calculated value for the beam size approaching the value expected from the optics for larger changes in $K_2$.
  }
   \label{fig:main_85_31}
\end{figure}

\paragraph{Discussion of Systematic Error in Beam Size Calculation}
Figure~\ref{fig:main_85_31}
shows the example of the contributions to the beam size calculation for the sextupole 10AW scan 85. This is the example $K_2$ scan used in the sections above. The optical functions in the CESR model predict a horizontal beam size of 1.09~mm at this sextupole. The contribution from $Y_0 \simeq -0.4$~mm is omitted because its contribution is small compared to that of $X_0$. A dependence on $\Delta K_2$ is observed, and results in a dependence of the calculated beam size on the magnitude of the sextupole strength change $\Delta K_2$. No such dependence is observed for $\Delta b_1 / \Delta K_2L$ ($X_0$).
The beam size calculated from the $K_2$ scan is observed become more precise and to approach the value expected from the optics for larger strength changes. 
The beam size dependence on $K_2$ arises from a contribution to the horizontal angle change which appears to be non-sextupole in nature.

\begin{figure}[htbp]
\centering
\includegraphics[width=\columnwidth]{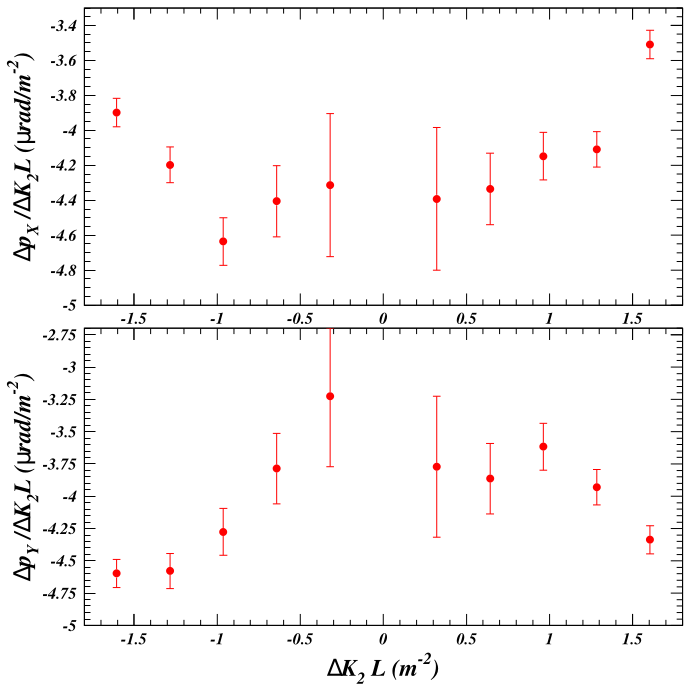}
\caption{The values for the vertical angle change slope show a dependence on $\Delta K_2$ opposite to that for the horizontal angle change slope. It is plausible that the vertical angle change could be used to correct the horizontal angle to correct for the apparently non-sextupole contribution.
  }
   \label{fig:main_85_32}
\end{figure}

Figure~\ref{fig:main_85_32}
shows that the vertical angle change has a similar, but opposite, dependence on the change in sextupole strength.
It has been shown for this case that an ad hoc assumption that the non-sextupole contribution to $\Delta p_X$ is half its contribution to  $\Delta p_Y$ reduces the difference between the calculated beam size and the value expected from the optics from 20~$\sigma$ to 1.7~$\sigma$~\cite{crittenden_25jul2023}, however without knowing the source of this non-sextupole contribution, such an assumption is not justifiable.

It is plausible that the non-sextupole contribution to $\Delta p_Y$
\begin{equation}
    \Delta p_Y^{\rm nonsext} = \Delta p_Y^{\rm meas} - \Delta K_2 L \; X_0 \; Y_0
\end{equation}
could be used to calculate the non-sextupole contribution to $\Delta p_X$
\begin{equation}
    \Delta p_X^{\rm nonsext} = \Delta p_X^{\rm meas} - \frac{1}{2} \Delta K_2 L \; (X_0^2 - Y_0^2).
\end{equation}
The curl relation
\begin{equation}
  \partial{x} \; \partial{B_X} = \partial{y} \; \partial{B_Y}
\end{equation}
may provide an answer to this puzzle since the horizontal and vertical angle changes along the trajectory are integrals of the corresponding field components.
If there is negligible contribution from fringe fields, we have in addition from the divergence equation
\begin{equation}
  \partial{y} \; \partial{B_X} = - \partial{x} \; \partial{B_Y}.
\end{equation}

A promising next step will be to use the measured anomalous horizontal and vertical deflections' dependence on the beam position in the sextupole to identify the multipole content of this non-sextupole contribution to the magnetic field.

\section{Conclusions}
\label{sec:conclusions}
\subsection{Sextupole Calibration Correction Factors}
The sextupole calibration procedure was updated in 2022 to use a custom closed-bump for each sextupole, and the fit to the tune change versus beam position was updated. During 2022 and 2023, 155 calibration data sets were recorded for the 76~sextupoles. A 3.1\% average correction was found and the RMS deviation of the correction factors is 12.5\%. The average uncertainty in the correction factors is found to be 1.7~\%. The RMS deviation in the uncertainties is 1.0~\%.

\subsection{Sextupole Alignment Values}
Horizontal and vertical misalignment values for 71 of the 76 sextupoles were measured using quadrupole and skew quadrupole terms derived from difference phase and coupling measurements, together with fits to the orbit data. The horizontal (vertical) misalignments average~-0.048~mm (-0.039~mm) and have an RMS spread of 1.0~mm (0.94~mm). The uncertainties in the misalignment determinations average~0.023~mm (0.021~mm) with an RMS spread of 0.027~mm (0.035~mm).

\subsection{Beam Size Calculations}
A measurement procedure and data analysis method has been developed which has sufficient statistical precision to determine the horizontal beam size at each CESR sextupole magnet with a precision of better than 30\%. However, an unknown systematic contribution to the uncertainty which is of magnitude comparable to the beam size itself spoils the results. The identification of this systematic contribution is the logical next step in the analysis. Success appears likely given the comprehensive nature of the data set.

\section{Acknowledgments}
The authors would like to acknowledge important critical input from the Cornell Electron-Ion-Collider/Energy Recovery research group throughout the course of this research project. Essential expert support for both the data-taking and the analysis was provided by the members of the CESR Operations Group: V.~Khachatryan, S.~Wang, J.Shanks, M.~Forster, and L.~Yang. The expertise of the CESR technical staff and operators were also essential to this work.
Helpful contributions were provided by participants in the CLASSE Research for Undergraduates program funded by the National Science Foundation: A.~Fagan, I.~Mishra and W.~Carbonell. Additional analysis support came from H.~Duan, J.Wang and A.~Shaked. We also acknowledge critical readings of the draft manuscript by Sophia Wang.
This work is supported by National Science Foundation award number DMR-1829070.

\pagebreak

\section*{References}

\raggedbottom

%

\end{document}